\documentclass[journal, letterpaper]{IEEEtran}
\usepackage{amsmath,amsthm,dsfont,amssymb}
\usepackage{tabularx,multirow}
\usepackage{graphicx,subfigure,comment}
\usepackage[table]{xcolor}
\usepackage{url} 
\usepackage{dsfont}
\usepackage{cite}
\usepackage{siunitx}
\usepackage{physics} 
\usepackage{pstool}
\usepackage{enumitem}
\usepackage{autonum} 
\usepackage{algorithm}
\usepackage[noend]{algpseudocode}
\graphicspath{{img/}}
\usepackage{soul}

\setlist[itemize]{leftmargin=*}
\setlist[enumerate]{leftmargin=*}

\newcommand{\Fig}[1]{Fig.~\ref{fig:#1}}
\newcommand{\Prope}[1]{Property~\ref{prope:#1}}
\newcommand{\Propo}[1]{Proposition~\ref{propo:#1}}

\newcommand{\Sec}[1]{Sec.~\ref{sec:#1}}

\newcommand{\Eq}[1]{(\ref{eq:#1})}
\newcommand{\Alg}[1]{Alg.~\ref{alg:#1}}
\newcommand{\Line}[1]{Line~\ref{line:#1}}

\newcommand{\ind}[1]{\mathds{1}_{#1}}
\newcommand{\Lc}{\mathcal{L}}
\newcommand{\Ac}{\mathcal{A}}

\newcommand{\Ic}{\mathcal{I}}
\newcommand{\Xc}{\mathcal{X}}
\newcommand{\Yc}{\mathcal{Y}}
\newcommand{\Sc}{\mathcal{S}}
\newcommand{\Tc}{\mathcal{T}}
\newcommand{\Uc}{\mathcal{U}}
\newcommand{\Pb}{\mathbf{P}}
\newcommand{\Qb}{\mathbf{Q}}
\newcommand{\NN}{\mathds{N}} 

\newtheorem{theorem}{Theorem}
\newtheorem{property}{Property}
\newtheorem{proposition}{Proposition}
\newtheorem{lemma}{Lemma}

\newtheorem{corollary}{Corollary}

\allowdisplaybreaks

\begin{document}

\title{
Network Support for High-performance\\ 
Distributed Machine Learning
}

\author{Francesco~Malandrino,~\IEEEmembership{Senior~Member,~IEEE,}
Carla~Fabiana~Chiasserini,~\IEEEmembership{Fellow,~IEEE,}
Nuria~Molner,~\IEEEmembership{Student~Member,~IEEE,}
and~Antonio~de~la~Oliva,~\IEEEmembership{Member,~IEEE}
\thanks{F.~Malandrino and C.~F.~Chiasserini are with CNR-IEIIT and CNIT, Italy. C.~F.~Chiasserini is with Politecnico di Torino, Italy. N.~Molner is with Universitat Polit\`ecnica de Val\`encia (iTEAM-UPV), Spain. A.~de~la~Oliva is with Universidad~Carlos~III~de~Madrid, Spain.}
\thanks{When this work was conducted, N.~Molner was with IMDEA Networks Institute and Universidad Carlos III de Madrid.}
\thanks{This work was supported through the EU Hexa-X project (Grant no. 101015956) and the NPRP-S 13th Cycle Grant No.\,NPRP13S-0205-200265 from the Qatar National Research Fund (a member of Qatar Foundation).  The views expressed are those of the authors and do not necessarily represent the projects.}
} 
\maketitle

\maketitle

\begin{abstract}
The traditional approach to distributed machine learning is to adapt
learning algorithms to the network, e.g., reducing updates to curb
overhead. Networks based on intelligent edge, instead, make it possible to
follow the opposite approach, i.e., to define the logical network topology 
 {\em around} the learning task to perform, so as to meet the
 desired  learning performance. 
In this paper, we  propose a system model that captures such aspects 
in the context of supervised machine learning, accounting 
for both learning nodes (that perform computations) and
information nodes (that provide data). We then formulate the 
problem of  selecting (i) which learning and  information
nodes should cooperate to complete the learning task, and (ii)
the number of epochs to run, in order
to minimize the learning cost while meeting the target prediction error
and execution time. 
After  proving important properties of the above problem, we  
devise an 
algorithm, named 
DoubleClimb, that can find  a $1+1/|\Ic|$-competitive solution (with 
$\Ic$ being the set of information nodes), with cubic {\em worst-case} complexity.
 Our performance evaluation, leveraging a real-world network topology 
and considering both classification and regression tasks, 
 also shows that DoubleClimb closely matches the optimum, outperforming state-of-the-art alternatives.
\end{abstract}

\section{Introduction}
\label{sec:intro}
Owing to the ever-increasing scale and complexity of the learning tasks to perform,
machine learning (ML) algorithms have swiftly been extended to work in a distributed
fashion, 
with the purpose of leveraging the computational capability of multiple nodes, 
possibly across multiple datacenters~\cite{li2014scaling,pham2018cooperative,distributedQlearning,8340193} 
and/or allowing nodes belonging to different parties to cooperate in a 
learning task without sharing sensitive data~\cite{wang2019adaptive,zhuo2019federated,konen2016federatedStrategies}.

More recently, distributed ML has  emerged also as an excellent match  for 
new generation (5G-and-beyond) networks. 
It can be used for the management of the network (as envisioned by such initiatives as ETSI ZSM~\cite{ETSI-ZSM}, 
ENI~\cite{ETSI-ENI}, and O-RAN~\cite{O-RAN}), as well as to enable user services within  
the so-called {\em intelligent edge}~\cite{xiao2020towards}.
In general, new generation networks can  (a) integrate a wide number
of {\em heterogeneous} nodes, including those
that can provide the data used for ML tasks, (b) provide a distributed computational infrastructure needed to run
the ML algorithms (see e.g., \cite{ETSI-36}), and (c) be dynamically reconfigured so as to
perform the ML task at hand with the required
performance.  

However, implementing an ML task in a  5G-and-beyond network also
poses important challenges. Specifically, it requires to define the {\em logical
 topology} of the nodes that cooperate towards the ML task, i.e., making decisions on:
\begin{itemize}
    \item which computing nodes in the different
      locations of the network edge should interact during the learning process;
    \item how many (and which) data sources to exploit, and which computing nodes
     should  receive their data.
\end{itemize}
The above decisions influence each other, often in counterintuitive
ways: as an example, seeking information from too many nodes may
result in longer learning times, due to the additional
waiting. Furthermore, a given target learning error (e.g., classification accuracy) may be reached through
alternative, completely different approaches, e.g., collecting a
significant quantity of information {\em or} performing more 
epochs to process a smaller set of data.

In spite of the wide usage of ML in mobile networks and the considerable attention devoted to it, 
most of the works aim at 
exploiting the network more efficiently, e.g., reducing
the overhead~\cite{li2014scaling,kadav2016asap} or dealing with
straggling nodes~\cite{li2018near}. Just a small number of
recent works~\cite{neglia,wang2019adaptive} have characterized the
impact of 
the network topology on the performance of distributed ML,
providing interesting insights on, e.g., the optimal network
connectivity.
However, {\em none} of these works tackle the
problem of  defining  the logical network  topology {\em
  around} the ML task to perform.

In this work, we focus on distributed, supervised learning, 
and aim at filling this gap by making the following main contributions:
\begin{itemize}
\item we  develop a
      system model that  can  represent several relevant supervised
      ML tasks and account for the specific features of a 
   5G-and-beyond environment, most notably, the interaction between learning nodes and information nodes;
    \item we formulate the problem of choosing the computing nodes and
      data sources, as well as the 
      links connecting them, with the goal of minimizing the (monetary or energy) cost 
    of  the  learning process, subject to prediction quality and learning time requirements;
    \item we prove that the problem is NP hard, but also, and most
      importantly, that it is submodular. In particular, although its
      constraints are 
      not monotonically increasing, we show that it can be
      solved via an iterative algorithm with
      excellent  competitive ratio guarantees;
    \item we propose an
      iterative algorithm, called DoubleClimb, which has cubic {\em worst-case} time
      complexity and  attains a {\em $1+1/|\Ic|$~competitive ratio}, with $\Ic$ being the set of
       information nodes. We evaluate DoubleClimb over a real-world topology, showing
      that it closely matches optimal decisions and substantially outperforms state-of-the-art alternatives.
\end{itemize}

The rest of the paper is organized as follows. After reviewing related
work in \Sec{relwork}, we describe our system model and how it
can represent different supervised ML tasks in \Sec{model}. In \Sec{problem}, we 
formulate  the problem we tackle and discuss its complexity. 
\Sec{characterize}  characterizes the  learning performance, 
while important properties of our problem are proven in \Sec{analysis}. 
We then present the DoubleClimb algorithm and
analyze its complexity in \Sec{algo}, before evaluating its
performance in \Sec{results}. 
We conclude the paper in \Sec{conclusion}.


\section{Related Work}
\label{sec:relwork}

Our work is related to the body of research works on distributed learning.
In this context, in the simplest scenarios~\cite{levine2020offline},
all training data is known before the training itself starts, 
and the purpose of performing distributed learning is simply to leverage more computational power. A more complex variation 
is represented by {\em active learning} where new information arrives
during the learning process, and is combined with the offline training
set~\cite{alaa,8600752}.
Applications include drone planning~\cite{pham2018cooperative}
and network management~\cite{chen2018communication,li2019accelerating}.

{\em Federated learning} is a more recent trend, tackling scenarios where
participating devices are not required to share potentially sensitive
data~\cite{konen2015federatedOptimization,
  konen2016federatedStrategies}. 
Depending upon the specific scenario, new data may or may not arrive during the training process. 

Several works propose generic methodologies to
mitigate common hurdles of distributed ML, including
scaling the parameter servers~\cite{li2014scaling}, dealing with
slower nodes~\cite{li2018near}, and trading learning efficiency for
convergence speed~\cite{kadav2016asap}. All these works propose novel
algorithms and/or approaches to {\em adapt to} the existing network
structure, e.g., by limiting the overhead, to perform the
learning task at hand as efficiently as possible. Importantly, {\em none} of them envision to do the opposite, i.e., adapting the nodes'
interaction  to the learning task.

Some works seek to theoretically characterize the convergence of supervised ML
and how it is influenced by the cooperation among learning
nodes. The study in \cite{8340193} 
characterizes the convergence of a wide class of multi-agent
algorithms. 
Using tools from spectral graph analysis, it establishes a relation between
the topology formed by pairs of cooperating nodes and the convergence
of the algorithm they run. \cite{neglia} focuses on 
distributed ML over regular topologies, and seeks to establish
the graph degree associated with the shortest convergence  {\em time}
-- as opposed to the lowest number of epochs --, finding that such
a degree depends on the distribution of the nodes' computing
time. Through similar steps and targeting a resource-constrained 
edge-computing scenario,
\cite{wang2019adaptive} searches for the optimal trade-off between
local computation and global parameter exchange in federated learning scenarios. 
With respect
to~\cite{8340193,neglia,wang2019adaptive}, we (i) seek to adapt the
logical network topology to the learning task, and (ii) consider not only learning nodes (in charge of processing information), but also information nodes, where data comes from. The latter is especially critical, as it allows us to characterize and study the trade-off between gathering information and extracting knowledge from it.

\begin{table}[tb]
    \caption{Main notation
        \label{tab:notation}
    } 
    \begin{center}
        {\renewcommand{\arraystretch}{1.2}%
            \begin{tabular}{|c|c|}
                \hline
                \textbf{Symbol} & {\textbf{Meaning}} \\
                \hline
                $\Lc$, $\Ic$ & L-nodes and I-nodes set (resp.)\\ 
                \hline
                $\rho_i(t)$ & pdf of  sample generation time at I-node $i\in \Ic$  \\
                \hline
                $r_i$ & ave. no. of samples per epoch by I-node $i$ \\
                \hline
                \multirow{2}{*}{$X^k_l$} & amount of samples at the beginning of epoch $k$ \\
                & at L-node $l$ \\
                \hline
                $c_l$, $c_i$ & operational cost of L-node $l$  and I-node $i$ (resp.)\\
                \hline
                $c_{l,l'}$ & communication cost between L-nodes $l,l'$ \\
                \hline 
                $c_{i,l}$ & communication cost between I-node $i$  and L-node $l$ \\
                \hline
                $\epsilon^{\max}$ & maximum learning error \\
                \hline
                $T^{\max}$ & maximum duration of the learning process \\
                \hline
                \hline
                \multirow{2}{*}{$p(l,l')$} & binary variable determining if L-nodes \\
                & $l$ and $l'$ cooperate (matrix $\Pb$) \\
                \hline
                \multirow{2}{*}{$q(i,l)$} & binary variable determining if L-node node $l$ obtains  \\
                & samples from I-node $i$ (matrix $\Qb$) \\ 
                \hline
                $K$ & number of epochs to run \\
                \hline
                $\tau^k_l(t)$ & pdf of the computation time at L-node $l$ and epoch $k$ \\
                \hline
                $\epsilon^K(\Pb,\Qb)$ &  global error at the end of the whole learning process \\
                \hline
                $T^K(\Pb,\Qb)$ & expected time to complete the whole  learning process \\
                \hline
                $C^K(\Pb,\Qb)$ & global cost for running the whole learning process \\
                \hline
        \end{tabular}}
    \end{center}
\end{table}

\section{System Model}
\label{sec:model}

Our system model addresses a generic distributed, supervised ML task 
where multiple nodes cooperatively seek to minimize a {\em loss function},
via gradient descent approaches such as 
the {\em stochastic gradient descent} (SGD) 
algorithm~\cite{neglia,wang2019adaptive,shamir2013stochastic,distributedQlearning}. 
In the following, we discuss how the behavior of individual nodes and their interactions 
are described by our system model, with reference to different real-world ML approaches.

{\bf Nodes' interactions.}
A {\em unique} feature of our model is its ability to 
capture the presence of 
two different types of nodes:
\begin{itemize}
    \item {\em learning nodes}, or L-nodes for short, that, having
      computational capabilities, run the ML algorithm and can
      exchange gradient data during  learning; we denote
      their set by  $\Lc$;
    \item {\em information nodes}, or I-nodes for short, which can
      provide information to the L-nodes; we denote
      their set by  $\Ic$.
\end{itemize}
Real-world counterparts of L-nodes include physical servers and
virtual machines running at the intelligent network edge~\cite{xiao2020towards} or in the cloud. I-nodes, on
the other hand, represent such entities as monitoring platforms,
 network nodes, and sensors. 

\begin{figure}[th]
\centering
    \def\svgwidth{\columnwidth}
  \begin{scriptsize}
  \textsf{
    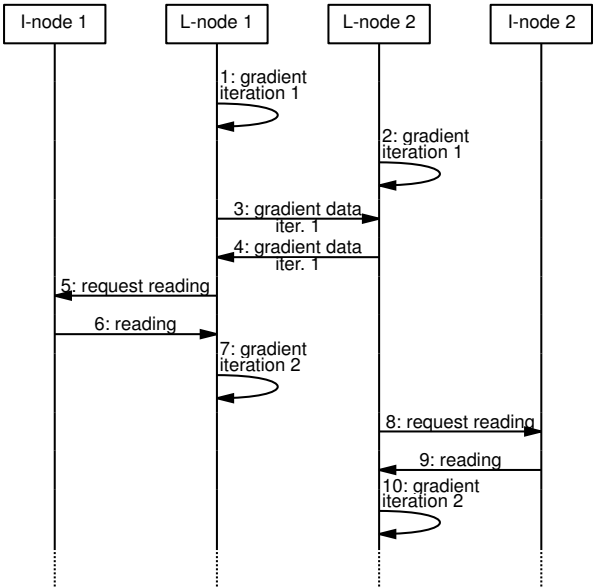
}
\end{scriptsize}
    \caption{
    Scheme of the interactions between L- and I-nodes in a
      general case.
      \label{fig:sequence}
    } 
\end{figure}

In our system model, L-nodes behave in a similar way to their
equivalents in~\cite{neglia,wang2019adaptive}. 
Their high-level goal is to cooperatively train a ML model network, and do so by minimizing a loss function 
via distributed optimization. 
The computation time at each epoch of the learning process at a generic node $l\in \Lc$ follows an arbitrary
distribution with probability density function
(pdf)~$\tau^k_l(t)$,
which also accounts for the node capability and the performance of the algorithm it runs.
Note that, in the most general case, such a pdf 
depends  on the current epoch ($k$) of the learning process, since the amount of samples used for learning may
vary from an epoch to the next one.
This reflects the need to exploit all the available data as soon as 
it becomes available~\cite{hard2018federated,8600752}, 
as opposed to training on a fixed number of samples as in more static scenarios. 
 L-nodes are logically connected to form an arbitrary {\em logical topology}, i.e., a graph where vertices represent L-nodes and edges,
hereinafter referred to as  L-L edges, represent 
the logical links connecting them.  
As exemplified in \Fig{sequence} (steps 3--4), after every epoch,
each L-node sends its gradient data to its neighboring  L-nodes  on the
logical topology, and waits for them to do the same before moving on. 
The logical topology, i.e., which pairs of L-nodes are neighbors and exchange gradient data, 
is one of our main decision variables. 

Each L-node can be logically connected to one or more I-nodes, through
the so-called  I-L edges. Only  I-nodes
that are connected to at least one L-node are added to the logical
topology. 
After each  epoch
of the learning process, an L-node
requests data from the I-nodes it is connected to (steps~5 and~8 in \Fig{sequence}),
receiving such data (steps~6 and~9 in \Fig{sequence}) after a {\em sample sending time} modeled by 
pdf~$\rho_i(t)$. In the following, we denote with~$r_i$~the expected number of samples provided by I-node~$i$ at each epoch.
The received samples are used
by an L-node $l$ to perform the next epoch, in addition to the
data it received in the previous epochs
and the number $X^0_l$ of (offline) samples initially available at
$l$. Note that this behavior is compatible with current, widely
deployed applications (e.g., IoT) using publish/subscribe mechanisms,
such as MQTT \cite{MQTT}, or Zenoh \cite{zenoh}, or even the notification mechanisms included
in the 3GPP Service Based Architecture \cite{TS23.501} of Release~15 and above.

Both L-nodes and  I-nodes have per-epoch operational costs, denoted by~$c_l$ and~$c_i$, 
respectively. Moreover, communication between nodes that are 
neighbors in the logical topology involve additional costs, denoted
by~$c_{l,l'}$ or~$c_{i,l}$ depending on the type of nodes. 
In general, such costs vary across different pairs of nodes, which also account 
for the fact that a logical link may correspond to multiple physical ones, hence, 
entail a higher cost due to energy and/or infrastructure payments.

{\bf Modeling real-world supervised ML tasks.}
As mentioned, our model can describe a wide range of real-world ML tasks, falling in the category of {\em supervised learning}, for which a ground truth is available. The most prominent examples of supervised learning tasks are classification (where the quantity to predict is discrete, e.g., whether or not a given transaction is fraudulent) and regression (where the quantity to predict is continuous).

In a distributed setting, supervised learning can be performed in two main modes:
\begin{itemize}
    \item  {\em distributed learning with static data}, where no new data arrive
      during the learning process. 
In this case, there are no I-nodes, and each L-node learns from its $X^0_l$~initial samples, as well as the gradient data from the other L-nodes;
    \item {\em active learning}~\cite{alaa}, where new samples can be
      collected from data sources (e.g., sensors) during the learning
      process so as to improve the learning
      quality. In this case, the network topology includes both L-
      and I-nodes. 
\end{itemize}
Importantly, our model can also capture {\em federated learning}~\cite{zhuo2019federated,yang2019federated,wang2019adaptive}, an emerging paradigm whereby different devices (e.g., smartphones) cooperatively train a model without sharing (potentially sensitive) data. In this case, each device is modeled as an L-node; if, in the specific scenario at hand, devices collect or generate additional information while learning, an I-node per device is added, only connected to the corresponding L-node.

For all tasks and approaches,  
our model can capture the cases where the communication between nodes happens in a peer-to-peer fashion~\cite{neglia,8340193}, as well as those when it is mediated by a {\em parameter server}, also known as {\em broker}~\cite{kadav2016asap,yang2019federated,wang2019adaptive}. In the latter case, the logical topology created by the L-nodes is fully connected.

\section{Problem Formulation and Approach}
\label{sec:problem}

Our decisions concern 
which nodes' interactions should be enabled, and the number of
epochs to execute during the learning process. We thus define
the following decision variables:
\begin{itemize}
    \item the set of binary variables~$p(l,l')\in\{0,1\}$,
      expressing whether L-nodes~$l$ and~$l'$ cooperate during 
      learning;
    \item the set of binary variables~$q(i,l)\in\{0,1\}$, expressing
      whether L-node~$l\in\Lc$ obtains samples from I-node~$i\in\Ic$;
\item the total number of epochs, $K$, to perform so that 
  the learning task meets  the desired 
  learning quality and execution time.
\end{itemize}
For compactness of notation, we will collect the $p$- and
$q$-variables in matrices $\Pb=\{p(l,l')\}$ and~$\Qb=\{q(i,l)\}$,
respectively. Given the
decisions~$\Pb$, $\Qb$, and $K$, we can compute the following system performance metrics:
\begin{itemize}
    \item the expected time required to the system to complete the
      learning process, denoted by~$T^K(\Pb,\Qb)$;
    \item the total cost $C^K(\Pb,\Qb)$, incurred by the system to complete the
      learning process,
which  accounts for (i) the cost of the infrastructure required to run the distributed learning, 
and (ii) the cost of the communication between the involved nodes;
    \item the (system-wide) learning error~$\epsilon^K(\Pb,\Qb)$ at
      the end of the learning process (i.e., after $K$  epochs).
\end{itemize}
Notice that the above error, cost, and learning time may depend upon other quantities, 
e.g., the number of samples available for training; however, to simplify the notation, 
we will write explicitly only the dependences on  our  
decision variables~$K$, $\Pb$ and~$\Qb$ as their indices.
Also, it is important to point out that in general the concrete definition of error~$\epsilon$  
depends on the type of learning task being performed, e.g.,
\begin{itemize}
    \item for classification tasks, $\epsilon\triangleq 1-\alpha$, 
    where $\alpha$ is the classification accuracy (i.e., the rate of 
    correctly labeled items);
    \item for regression tasks, $\epsilon\triangleq 1-R^2$, 
    where $R^2$ is the coefficient of determination~\cite{nagelkerke1991note}.
\end{itemize}
In both cases, $\epsilon=0$ corresponds to perfect learning, while larger  $\epsilon$ values 
identify worse learning quality, i.e., higher error. 
In the remainder of the paper, we use {\em learning error} or {\em learning quality} when 
referring to generic machine learning, and more precise terms (e.g., {\em accuracy} for classification) 
when discussing specific learning tasks.

Our objective is to minimize the total cost, while ensuring that  the final learning
error does not exceed the
limit~$\epsilon^{\max}$, i.e., $\epsilon^K(\Pb,\Qb) \leq
\epsilon^{\max}$, and the learning is completed within the target time, i.e., $ T^K(\Pb,\Qb) \leq
T^{\max}$. The problem can then be synthetically formulated as:
\begin{equation}
\label{eq:obj}
\min_{\Pb,\Qb,K} C^K(\Pb,\Qb),
\end{equation}
\begin{equation}
\label{eq:constr}
\text{s.t.} \min\left\{\frac{\epsilon^{\max}}{\epsilon^K(\Pb,\Qb)},\frac{T^{\max}}{T^K(\Pb,\Qb)}\right\}\geq 1.
\end{equation}
The  problem is combinatorial in nature and includes a large number of
binary variables (the elements of matrices $\Pb$ and $\Qb$). This makes it very hard  to
solve, even without considering the complexity of
computing the quantities $C^K(\Pb,\Qb)$, $\epsilon^K(\Pb,\Qb)$, and $T^K(\Pb,\Qb)$.  
Specifically, we prove in \Sec{analysis} that the problem is NP hard.

Remarkably, in spite of the problem
complexity, we can design an efficient
and provably effective 
solution strategy.  We do so by first characterizing the system
performance as functions of the problem decision variables 
(\Sec{characterize}), and then showing that the problem in
\Eq{obj} and \Eq{constr} is {\em submodular}  (\Sec{analysis}). 
Leveraging this result, we can devise the DoubleClimb algorithm
(\Sec{algo}), which has cubic worst-case  time complexity and 
proves to be $1+1/|\Ic|$ competitive.

\section{Characterizing the Behavior of the Learning Process}
\label{sec:characterize}

In order to make our decisions, i.e., to choose the best values for the~$\Pb$ and~$\Qb$ matrices, 
we need to understand their impact on the learning behavior, e.g., how the learning quality evolves 
across epochs. In spite of its importance, and the vast quantity of research devoted to it, 
the goal of fully characterizing a learning process has not yet been achieved. 
Indeed, as reported in~\cite{hestness2017deep}, the learning process can best be described 
as {\em empirically predictable}. In other words, (i) learning tasks consistently behave 
according to the same laws, but (ii) the parameters of such laws depend upon the concrete 
learning task at hand (e.g., the selected neural network architecture and the data used for training).
In this section, we describe how to characterize the learning accuracy (\Sec{sub-accuracy}), 
the time it takes (\Sec{sub-time}), and the associated cost (\Sec{sub-cost}).

\subsection{Learning accuracy}
\label{sec:sub-accuracy}

One of the main metrics in our problem
is the learning quality,  
or, equivalently, error $\epsilon$, 
and how it changes according to (i) the number of epochs being performed, (ii) 
the connectivity among L-nodes, and (iii) the connectivity between I- and L-nodes. 
Concerning the first two aspects, \cite{8340193,neglia}~have derived a  
square-root behavior, which can be expressed as:
\begin{equation}
\nonumber
\epsilon^K=a_1+\frac{a_2}{\sqrt{K\gamma}}
\end{equation}
where $K$~is the number of epochs performed, and $\gamma$~is the spectral 
gap\footnote{The spectral gap of a graph is the difference between the moduli of the two 
largest eigenvalues of its adjacency matrix.} of the graph describing the cooperation among L-nodes. 
Notice that such a result has been proven without reference to a specific dataset or neural 
network architecture; these elements are accounted for through the $a_1$ and $a_2$ coefficients. 

Then let us define $X$ as the number of available samples, averaged over epochs and learning nodes. 
The relationship between the average size~$X$ of local datasets and the learning quality 
is a case of ``empirical predictability'': in spite of the lack of theoretical results 
explaining such a behavior, all measurement works we have 
surveyed~\cite{sun2017revisiting,linjordet2019impact,alaa,perlich2003tree,distributedQlearning}, 
as well as our own experiments, have invariably found a logarithmic law, i.e.,
\begin{equation}
\nonumber
\epsilon^K\propto\log\left(a_3+X\right)\,.
\end{equation}

Combining the two above expressions, we can write: 
\begin{equation}
\label{eq:generic-law}
\epsilon^K=c_1+\frac{c_2\log(c_3+X)}{\sqrt{K\gamma}}\,.
\end{equation}
In terms of our decision variables~$\Pb$ and~$\Qb$, $\gamma$ is the difference 
between the first and second eigenvalues of matrix~$\Pb$, i.e.,
\begin{equation}
\nonumber
\gamma=|\text{eig}_1(\Pb)|-|\text{eig}_2(\Pb)|,
\end{equation}
while the size~$X$ of local datasets can be written as:
\begin{equation}
\nonumber
X=\frac{1}{K|\Lc|}\sum_{l\in\Lc}\sum_{k=1}^K\left(X^0_l+\sum_{i\in\Ic}kr_iq(i,l)\right)\,,
\end{equation}
where $q(i,l)\in\{0,1\}$ is the element of~$\Qb$ describing whether I-node~$i$ is connected with L-node~$l$.
Notice that, by using expected values, we are able to write \Eq{generic-law} using deterministic, known quantities, in spite of the fact that the underlying process is stochastic in nature. 

The generic law in\,\Eq{generic-law} describes, as confirmed by overwhelming evidence~\cite{8340193,neglia,wang2019adaptive,hestness2017deep,sun2017revisiting,linjordet2019impact,alaa,perlich2003tree,distributedQlearning} 
 a very wide set of ML tasks in a very large set of applications. 
However, the concrete values of coefficients~$c_1$--$c_3$ depend upon the concrete learning task 
at hand, including the DNN architecture and dataset being used.
As a consequence, a small-scale {\em profiling} of the selected DNN and dataset is necessary, 
in order to establish the~$c_1$--$c_3$ coefficients; afterwards, our system model and the 
solution strategy described in \Sec{algo} can be leveraged to optimize the actual, 
large-scale learning. Such an approach has been used in~\cite{hestness2017deep}, and successfully 
validated over multiple learning tasks, models, datasets, and applications, 
including speech recognition using LSTM networks, image classification with convolutional 
networks, and human attention 
model with recurrent networks.

Importantly, once the concrete learning task to perform is known and the profiling phase 
has been completed, the exact values of all the quantities needed to compute \Eq{generic-law} 
are known, i.e., such values are known parameters of our problem (as opposed to random variables).

\subsection{Learning time}
\label{sec:sub-time}
We now consider that  the total number of epochs $K$, the pdfs~$\rho_i(t)$ of the
sample generation time at each I-node~$i$, and the pdfs~$\tau^k_l(t)$
of the computation time
of each L-node~$l$ at epoch $k$ are given. Recall that
$\tau^k_l(t)$ depends on $k$, as the presence of I-nodes in our system
model
implies that the computation time distribution must account for the
quantity~$X^k_l$ of available data at L-node $l$ and epoch $k$.
In view of the fact that the computation time of DNNs grows linearly~\cite{serpen2014complexity} with the quantity of data, we can write:
\begin{equation}
\label{eq:tau-k}
\tau_l^k(t)=\frac{X^k_l}{X^0}\tau_l^0(t)\,.
\end{equation}
Notice how the linear relationship in \Eq{tau-k} is consistent with real-world 
measurements~\cite{hestness2017deep}, theoretical studies~\cite{neglia,wang2019adaptive}, 
and the intuition that, especially when data is processed in (mini-) batches,  
processing twice the data requires twice the effort.

Also, we define the sets~$\Ic_l=\{i\in\Ic\colon q(i,l)=1\}$
and~$\Lc_l=\{l'\in\Lc\colon p(l,l')=1\}$ of I-nodes and L-nodes (resp.)
each L-node is connected with. 
Our goal is to compute~$T^K(\Pb,\Qb)$, i.e., the total time
required to complete the whole learning process.

As highlighted in \Fig{sequence}, at every epoch each L-node must perform the following steps:
\begin{itemize}
    \item wait for the information coming from the I-nodes~$i\in\Ic_l$;
    \item perform its own gradient computation;
    \item wait for the gradient data coming from the other
      L-nodes~$l'\in\Lc_l$ it is cooperating with.
\end{itemize}

The first step is complete when {\em all} nodes in~$\Ic_l$ send their
samples. Recalling that each I-node has a sample generation time
distributed with pdf~$\rho_i(t)$, we can derive  the cumulative
distribution function (CDF) of the maximum of a set of independent random
variables as the product of individual CDFs $R_i(t)$, i.e., $\prod_{i\in\Ic_l}R_i(t)$.
Once all data arrive, $l$ can perform its own gradient computation,
whose duration is distributed according to
pdf~$\tau^k_l(t)$. Recalling that the pdf of the sum of two independent random variables is the convolution of individual pdfs, we can write:
  $  h^k_l(t)= \tau^k_l(t) \ast \dv{(\prod_{i\in\Ic_l}R_i(t))}{t}$.

For the system as a whole to move to the next epoch, all L-nodes
must have received the gradient data they need. This, in turn,
requires the slowest L-node to have obtained its information and have 
performed the computation. Working again with CDFs, the time taken by such a node is distributed according to:
   $ H^k(t)=\prod_{l\in\Lc}H_l^k(t)$,
where $H_l^k(t)$ denotes the CDF of the time to complete  epoch $k$
at L-node $l$. By letting  $h^k(t)=\dv{H^k(t)}{t}$, the expected duration of
the learning process is then 
given by:  
\begin{equation}
    T^K(\Pb,\Qb)= \sum_{k=1}^K\int_0^\infty 
    x h^k(t)  dt.
\end{equation}

\begin{figure}[th]
\centering
\includegraphics[width=.4\textwidth]{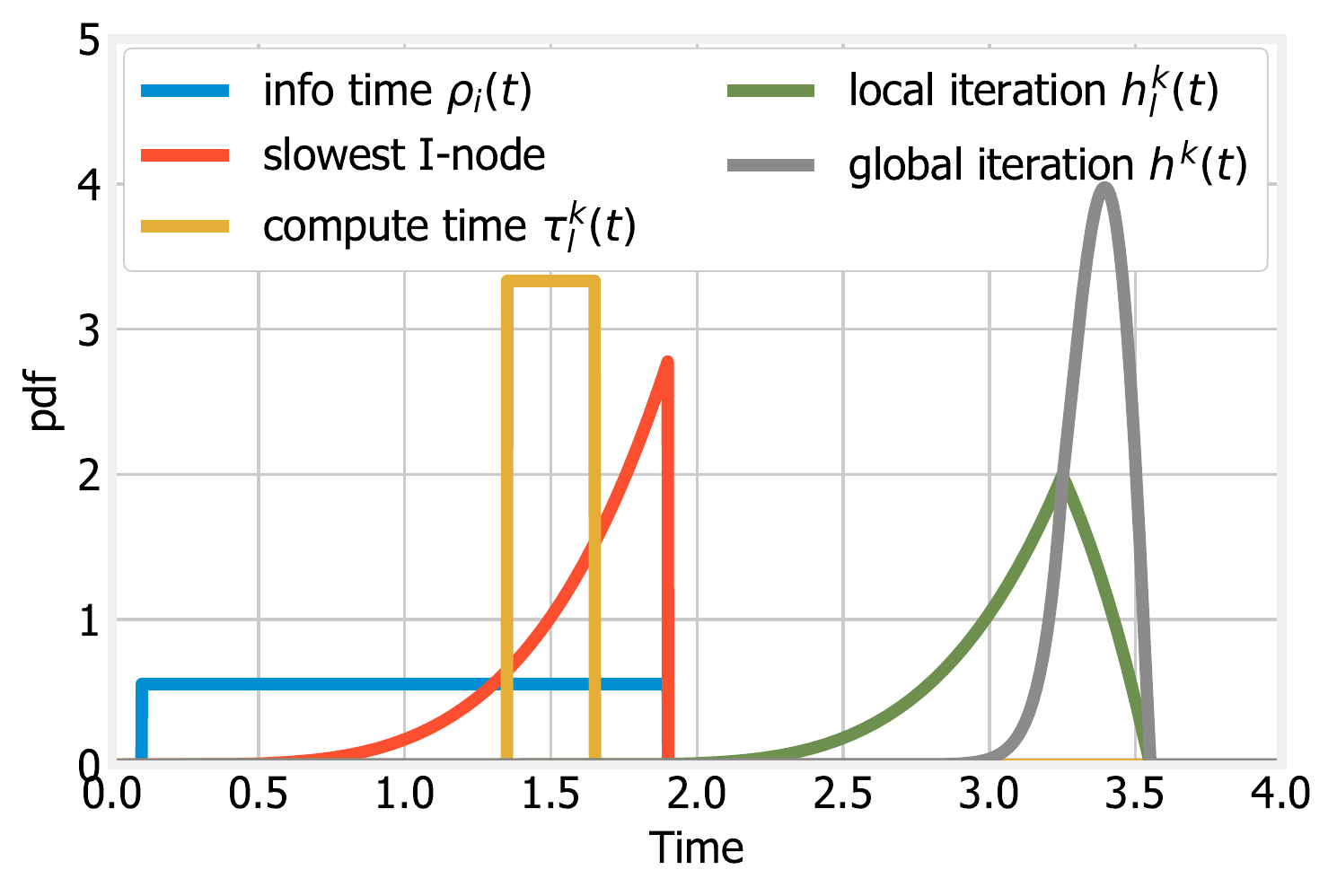}
\caption{
Toy scenario with~$|\Lc|=10$ and~$|\Ic|=5$ where both I-node sample generation times and 
L-node computation times are uniformly distributed. Left: pdfs of the I-node generation 
time~$\rho_i(t)$ (blue), of the time required by the slowest I-node (red) and 
of the compute time~$\tau^k_l(t)$ (yellow). Right: pdfs of the time taken by local (green) 
and global (gray) epochs.
    \label{fig:pdfs}
} 
\end{figure}

{\bf A numerical example.}
\Fig{pdfs} exemplifies our methodology in a case where both the I-node sample  
generation times and the L-node computation times are uniformly distributed; 
specifically,~$\rho_i(t)\sim\Uc(0.1,1.9)$ and~$\tau_l^k(t)\sim\Uc(1.35,1.65)$. 
Furthermore, there are~$|\Lc|=10$ L-nodes, each connected to~$|\Ic|=5$ I-nodes.

We begin from the blue line in the plot, representing~$\rho_i(t)$. 
To obtain the pdf of the sample generation time of the slowest I-node, 
we have to integrate~$\rho_i(t)$ (obtaining $R_i(t)$, a ramp-like function), 
then raise it to the $|\Ic|$-th power (obtaining a 5th-degree polynomial), 
and finally derive it, obtaining the 4th-degree polynomial shown by the red line in \Fig{pdfs}.

We next perform the convolution between the latter pdf and~$\tau^k_l(t)$, 
represented by the yellow line in the plot. The result is~$h^k_l(t)$, 
represented by the green line in \Fig{pdfs}. 
The last step consists in computing the distribution of the time taken 
by the whole learning epoch, hence, by the slowest L-node. 
Integrating~$h^k_l(t)$, we obtain~$H^k_l(t)$, which we raise to the~$|\Lc|=10$-th power, 
and then derive it, obtaining the pdf~$h^k(t)$ shown by the gray curve in \Fig{pdfs}.

\begin{figure}[th]
	\centering
	\includegraphics[width=\columnwidth]{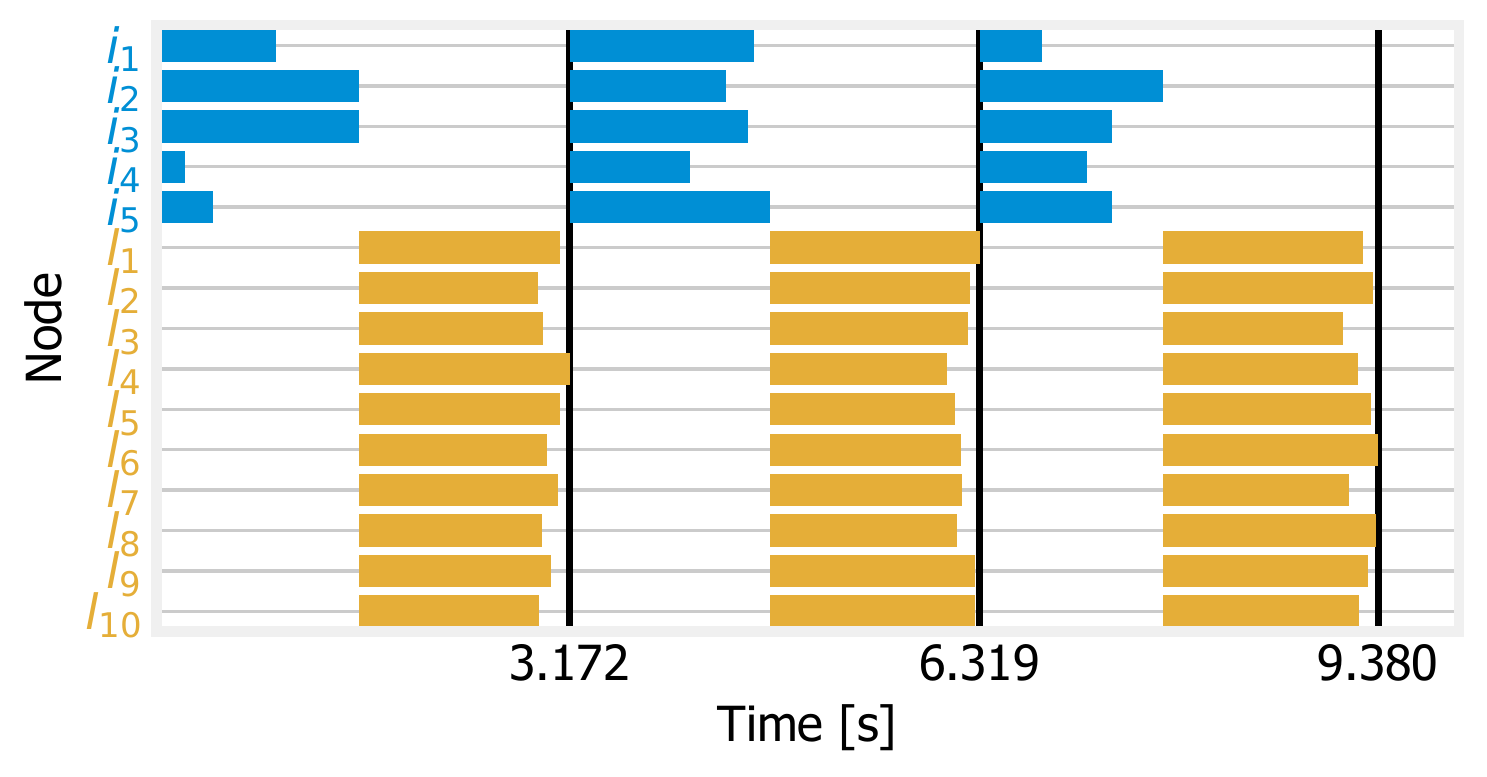}
	\includegraphics[width=\columnwidth]{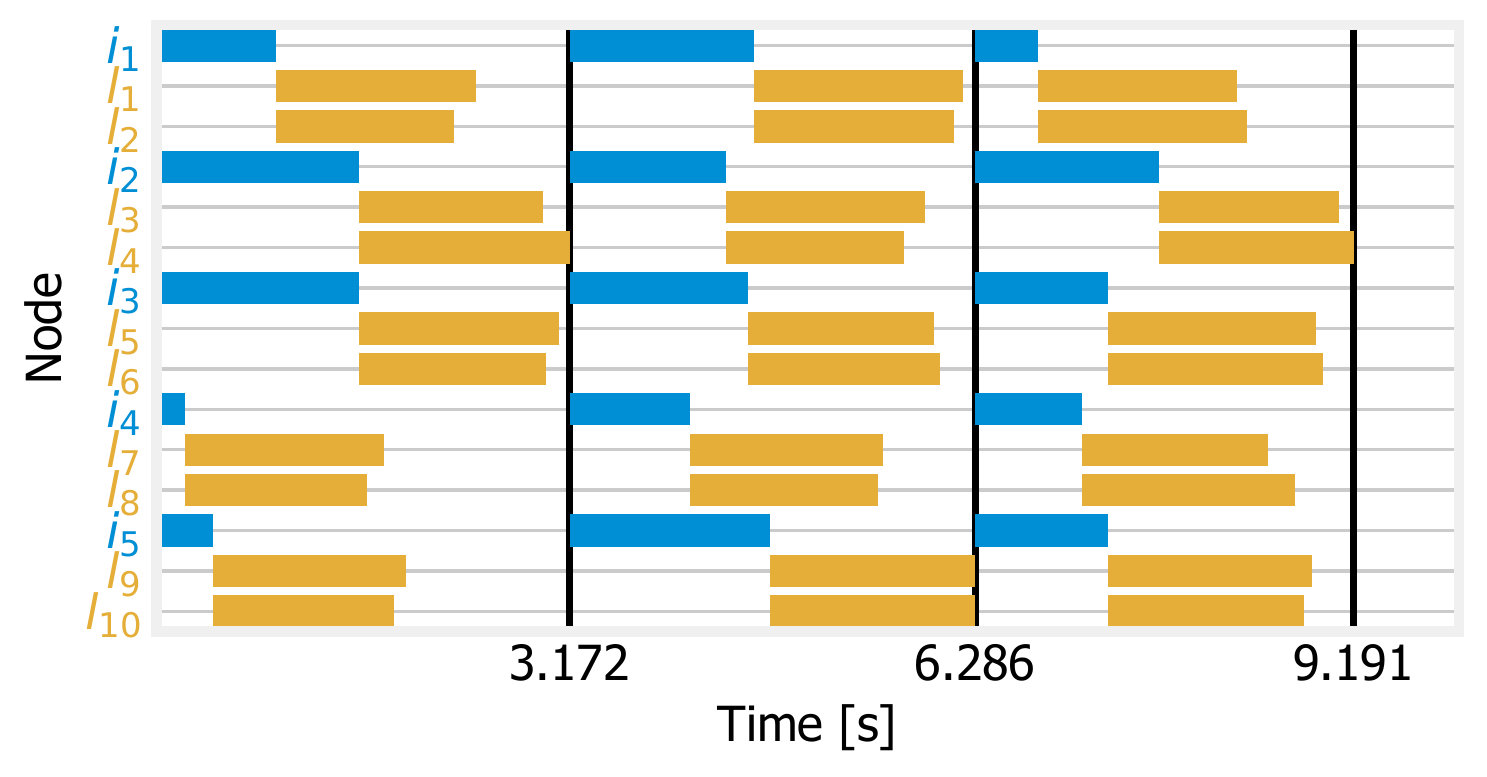}
	
	\caption{
			Toy scenario with~$|\Lc|=10$ and~$|\Ic|=5$ where both I-node sample generation times and 
			L-node computation times are uniformly distributed. Gantt charts show the activity of 
			I- and L-nodes (blue and yellow bars, respectively), when all L-nodes use all I-nodes (top) 
			and when each L-node only uses one I-node (bottom). Black vertical lines mark the end time of each epoch.
		\label{fig:gantt}
	} 
\end{figure}

\Fig{gantt} presents two Gantt charts showing the activity of I- and L-nodes 
(blue and yellow bars, respectively) over three epochs. The top plot refers to the case where 
all L-nodes use all I-nodes, as in \Fig{pdfs}: accordingly, it is possible to observe 
how all yellow bars start at the same time, after all blue bars finish. In the bottom plot, 
we move to a scenario where each L-node uses only one I-node 
(specifically, $l_1$ and~$l_2$ use~$i_1$, $l_3$ and~$l_4$ use~$i_2$, etc.). 
This allows many L-nodes to begin their work early, however, as we can see from the vertical bars, 
the overall decrease in epoch duration is modest -- even absent in the first epoch, 
where the slowest L-node had to wait for the slowest I-node. This is due to the fact that, 
in this toy example, both the I-node sample generation times and the L-node computation times 
are uniformly distributed: indeed, as also \cite{neglia}~reports, pruning I-L links is most 
beneficial when computation and generation times follow more skewed distributions.
At last, we note that the limited reduction in the learning time shown in the figure is due to the small size of the scenario;
nonetheless,  such a gain validates our approach.

{\bf Closed-form expression for special cases.}
	The methodology outlined above does not require any assumption on
	the~$\tau^k_l(t)$ and~$\rho_i(t)$ distributions, nor on the logical
	links between nodes, and the computations it requires can always  be performed
	numerically.  However, closed-form expressions are available in relevant special cases.
Let us focus on a scenario where (i) all nodes are connected to each other, and (ii) the computation
    and the sample generation times are i.i.d. and exponentially distributed with 
    parameter~$\lambda_{L}^{k}$ and $\lambda_I$, respectively. 
    Such a scenario is sufficiently simple to result in manageable expressions, but also sufficiently complex to allow us to properly illustrate the power of our methodology.

The computation time~$T^K$ can be written as:
	\begin{equation}
	    \label{eq:avgTimeScenario-exp}
	T^K{=}{-}\sum_{k=1}^{K} \sum_{\substack{\Ac\subset\NN \colon \\
	 |\Ac|=|\Ic|+2 \\ 
	 \sum_{a\in \Ac} a =|\Lc|}}\hspace{-2mm}
	\binom{|\Lc|}{\Ac}\frac{\prod_{w=1}^{|\Ic|+2}(A^k(\Ac,w))^{a_w}}{\lambda_I\sum_{w=1}^{|\Ic|}wa_w+\lambda_L^k a_{|\Ic|+2}}.
	\end{equation}
	In the above expression, the sum over $k$ accounts for all epoch, $k=1,\ldots, K$.
	The inner sum comes from the multinomial
	expansion \cite{multinomial} of a sum of 
	$|\Ic|+2$ terms (one for each I-node, one for the L-node connected
	to them, and one representing the coefficient) raised to the $|\Lc|$-th power, where each term is a polynomial  
	(see also the expression of $h^k_l(t)$).  
	Therefore, the inner summation is over all sets~$\Ac$ of natural numbers such
	that their size is~$|\Ic|+2$ and their sum is~$|\Lc|$, and 
	  $\binom{|\Lc|}{\Ac}=\frac{|\Lc|!}{\prod_{a\in\Ac}a!}$ is the
	multinomial coefficient. The term~$A^k(\Ac,w)$ associated with the
	$w$-th element of each set~$\Ac$ is: 
	\begin{equation}
	\nonumber
	A^k(\Ac,w)=
	\begin{cases}
	\sum_{z=1}^{|\Ic|}\binom{|\Ic|}{z}(-1)^{z{+}1}, &\text{if }w{=} |\Ic|+1\\
	\sum_{z=1}^{|\Ic|}\binom{|\Ic|}{z}(-1)^{z{+}1}\frac{z\lambda_I}{\lambda_L^k-w\lambda_I}, 
	&\text{if }w{=}|\Ac|\\
	\binom{|\Ic|}{w}(-1)^{w{+}1}\frac{\lambda_L^k}{w\lambda_I-\lambda_L^k}, &\text{otherwise.}
	\end{cases}
	\end{equation}
 
	A closed-form expression for the expected duration 
	of the learning process can also be obtained when  each  
	L-node receives information from all I-nodes, and  the I-nodes' sample generation times 
	 and the L-nodes's computation times are i.i.d. and uniformly distributed 
	over $(a_{I},b_{I})$ and $(a_{L}^{k},b_{L}^{k})$, respectively.
	For simplicity and without loss of generality, let us assume 
	 $a_{L}^{k} \leq a_{I} \leq b_{I} \leq b_{L}^{k}, \forall k$; then, we have: 
	\begin{align}
		\label{eq:avgTimeScenario-unif}
		&T^K{=}\sum_{k=1}^{K} \sum_{\substack{\Ac\subset\NN \colon \\
				|\Ac|=|\Ic|+2 \\ 
				\sum_{a\in \Ac} a =|\Lc|}}\hspace{-2mm}
		\binom{|\Lc|}{\Ac}
		\frac{\sum_{w=1}^{|\Ic|+1}wa_w}{\sum_{w=1}^{|\Ic|+1}wa_w + 1}{\times} \\
		\nonumber
		&{\times}\Bigg[\hspace{-1.3mm}\prod_{w{=}1}^{|\Ic|{+}2}\hspace{-2mm}(A_1^k(\Ac,w))^{a_w} 
		\Big(\hspace{-1mm}Z_1^{\sum\limits_{w{=}1}^{|\Ic|{+}1}wa_w{+}1}
		\hspace{-1mm}{-}Z_2^{\sum\limits_{w{=}1}^{|\Ic|{+}1}wa_w{+}1}\Big) \\
		\nonumber
		&{+}\prod_{w{=}1}^{|\Ic|{+}2}\hspace{-2mm}(A_2^k(\Ac,w))^{a_w} 
		\Big(\hspace{-1mm}Z_3^{\sum\limits_{w{=}1}^{|\Ic|{+}1}wa_w{+}1}
		\hspace{-1mm}{-}Z_4^{\sum\limits_{w{=}1}^{|\Ic|{+}1}wa_w{+}1}\Big) \\
		\nonumber
		&{+}\prod_{w{=}1}^{|\Ic|{+}2}\hspace{-2mm}(A_3^k(\Ac,w))^{a_w} 
		\Big(\hspace{-1mm}Z_5^{\sum\limits_{w{=}1}^{|\Ic|{+}1}wa_w{+}1}
		\hspace{-1mm}{-}Z_6^{\sum\limits_{w{=}1}^{|\Ic|{+}1}wa_w{+}1}\Big) 
		\hspace{-1.3mm}\Bigg] 
	\end{align}
	where $Z_1=a_{L}^{k}{+}b_{I}$, $Z_2=a_{L}^{k}{+}a_{I}$, $Z_3=b_{L}^{k}{+}a_{I}$, 
	$Z_4=a_{L}^{k}{+}b_{I}$, $Z_5=b_{L}^{k}{+}b_{I}$, $Z_6=b_{L}^{k}{+}a_{I}$.
	As in the previous case, the above expression comes from the multinomial expansion 
	\cite{multinomial}, and, after some algebra, one can obtain   
	the terms~$A_1^k(\Ac,w)$,~$A_2^k(\Ac,w)$, and~$A_3^k(\Ac,w)$ associated with the $w$-th 
	element of each set~$\Ac$, as: 
	\begin{equation}
		\nonumber
		A_1^k(\Ac,w){=}\hspace{-1mm}
		\begin{cases}
			\frac{(-2a_{I})^{|\Ic|}{-}\left(a_{L}^{k}{-}a_{I}\right)^{|\Ic|}}{\left(b_{I}{-}
			a_{I}\right)^{|\Ic|}(b_{L}^{k}{-}a_{L}^{k})}\hspace{-0.5mm}, &\hspace{-3mm}
			\text{if }w{=} 1\vspace{3mm}\\
			\frac{\left(a_{L}^{k}{-}a_{I}\right)^{|\Ic|}(|\Ic| (a_{L}^{k}{+}a_{I}){+}2a_{I}){+}
			({-}2a_{I})^{|\Ic|{+}1}}{(|\Ic|{+}1)\left(b_{I}{-}a_{I}\right)^{|\Ic|}
			(b_{L}^{k}{-}a_{L}^{k})}\hspace{-0.5mm}, &\hspace{-3mm}\text{if }w{=} |\Ac|
			\vspace{3mm}\\
			\frac{\binom{|\Ic|{+}1}{w} ({-}2a_{I})^{|\Ic|{+}1{-}w}}{(|\Ic|{+}1)
			\left(b_{I}{-}a_{I}\right)^{|\Ic|}(b_{L}^{k}{-}a_{L}^{k})}\hspace{-0.5mm}, 
			&\hspace{-3mm}\text{else.}
		\end{cases}
	\end{equation}
	\begin{equation}
		\nonumber
		A_2^k(\Ac,w){=}\hspace{-1mm}
		\begin{cases}
			A_1^k(\Ac,|\Ac|){+}\sum\limits_{z=1}^{|\Ic|{+}1} A_1^k(\Ac,z)(a_{L}^{k}{+}
			b_{I})^z{+} & \\ 
			{+}\frac{\left(a_{L}^{k}{-}a_{I}\right)^{|\Ic|{+}1}{-}\left(a_{L}^{k}{+}b_{I}{-}2a_{I}
			\right)^{|\Ic|{+}1}}{(|\Ic|{+}1)\left(b_{I}{-}a_{I}\right)^{|\Ic|}(b_{L}^{k}{-}
			a_{L}^{k})}{+} & \\
			{+}\frac{\left(b_{I}{+}a_{I}\right)^{|\Ic|{+}1}{-}(2a_{I})^{|\Ic|{+}1}}
			{(|\Ic|{+}1)\left(b_{I}{-}a_{I}\right)^{|\Ic|}(b_{L}^{k}{-}a_{L}^{k})}\hspace{-0.5mm}, 
			&\hspace{-8mm}\text{if }w{=} |\Ac|\vspace{3mm}\\
			\frac{{-}\binom{|\Ic|{+}1}{w} (\left({-}b_{I}{-}a_{I}\right)^{|\Ic|{+}1{-}w}
			{-}({-}2a_{I})^{|\Ic|{+}1{-}w})}{(|\Ic|{+}1)\left(b_{I}{-}a_{I}\right)^{|\Ic|}
			(b_{L}^{k}{-}a_{L}^{k})}\hspace{-0.5mm}, &\hspace{-2mm}\text{else.}
		\end{cases}
	\end{equation}
	\begin{equation}
		\nonumber
		A_3^k(\Ac,w)=\hspace{-1mm}
		\begin{cases}
			\frac{(b_{L}^{k}{-}a_{I})^{|\Ic|}{-}\left(b_{I}{+}a_{I}\right)^{|\Ic|}({-}1)^{|\Ic|}}
			{\left(b_{I}{-}a_{I}\right)^{|\Ic|}(b_{L}^{k}{-}a_{L}^{k})}\hspace{-0.5mm}, 
			&\hspace{-5mm}\text{if }w{=} 1\vspace{3mm}\\
			A_2^k(\Ac,|\Ac|){+}\sum\limits_{z=1}^{|\Ic|+1} A_2^k(\Ac,z)(b_{L}^{k}{+}a_{I})^z{-} 
			& \\
			{-}\frac{(|\Ic|{+}1)\left(b_{L}^{k}{-}a_{I}\right)^{|\Ic|}(b_{L}^{k}{+}a_{I})}{(|\Ic|
			{+}1)\left(b_{I}{-}a_{I}\right)^{|\Ic|}(b_{L}^{k}{-}a_{L}^{k})}{+} & \\
			{+}\frac{\left({-}b_{L}^{k}{-}a_{I}\right)^{|\Ic|{+}1}{-}
			\left(b_{L}^{k}{-}b_{I}\right)^{|\Ic|{+}1}}
			{(|\Ic|{+}1)\left(b_{I}{-}a_{I}\right)^{|\Ic|}(b_{L}^{k}{-}a_{L}^{k})}\hspace{-0.5mm}, 
			&\hspace{-7mm}\text{if }w{=} |\Ac|\vspace{3mm}\\
			\frac{\binom{|\Ic|{+}1}{w} \left({-}b_{I}{-}
			a_{I}\right)^{|\Ic|{+}1{-}w}}{(|\Ic|{+}1)\left(b_{I}{-}a_{I}\right)^{|\Ic|}
			(b_{L}^{k}{-}a_{L}^{k})}\hspace{-0.5mm}, &\hspace{-2mm}\text{else.}
		\end{cases}
	\end{equation}
	
	Intuitively, the three different terms~$A_{\ast}^k(\Ac,w)$ are 
	due to the convolution of the pdfs, which results in a piece-wise function 
	(see also the expression of $h^k_l(t)$). 
	The support of the different pieces of the function are as follows: 
	$\left[a_{L}^{k}+a_{I}, a_{L}^{k}+b_{I}\right)$ for the first piece where only one pdf 
	is active, $\left[a_{L}^{k}+b_{I}, b_{L}^{k}+a_{I}\right]$ for the second piece where 
	both pdfs are active and overlap, and $\left(b_{L}^{k}+a_{I}, b_{L}^{k}+b_{I}\right]$ 
	for the third piece where only the other pdf is active.

\subsection{Learning cost}
\label{sec:sub-cost}

We define the per-epoch cost as the sum of operational and
communication costs of  the L- and I- nodes
contributing to each epoch, i.e., 
\begin{eqnarray}
C(\Pb,\Qb)&\mathord{=}&\sum_{l\in \Lc} \left ( c_l \mathord{+} \sum_{l' \in \Lc}  c_{l,l'}
  p(l,l') \mathord{+} \sum_{i\in \Ic} c_{i,l} q(i,l) \right ) \nonumber\\
  \label{eq:cost}
&&\mathord{+} \sum_{i\in\Ic} 
c_i \mathds{1}_{\exists q(i,l)>0} \,. 
\end{eqnarray}
Then, we can write  the total learning cost over the $K$ epochs as 
$C^K(\Pb,\Qb)=K\cdot C(\Pb,\Qb)$.

\subsection{Number of epochs}
The number~$K$ of epochs needed to reach the target
error~$\epsilon^{\max}$ depends on two factors. The first is the
quantity of available training data: the more data
is available, the more the learning quality improves at each epoch. The
second is the level of cooperation between L-nodes: the more nodes
cooperate, the higher the quality achieved at each epoch. 
From \Eq{generic-law}, we get:
\begin{equation}
\nonumber
K\propto \frac{\log^2 X}{\gamma\left(\epsilon^{\max}\right)^2}.
\end{equation}
On the one hand, a high degree of
L-nodes  makes the learning process faster, as convergence requires fewer epochs; on the other hand, each epoch takes longer to complete as there are more nodes to wait for.

\section{Problem Analysis}
\label{sec:analysis}

We first prove that the problem at hand, formulated in \Sec{problem}, 
is NP hard. On the positive side, we also show that the problem objective function  
is submodular and non-decreasing, while the constraint is submodular and exhibits 
only one maximum 
(we prove the latter part separately for I-L and L-L edges). 
\begin{lemma}
The problem of optimally configuring the system for an ML task, 
expressed in \Eq{obj} and \Eq{constr}, is NP hard.
\end{lemma}
\begin{IEEEproof}
The proof  can be obtained via a reduction from the knapsack
problem~\cite{woeginger2003exact}, 
a combinatorial optimization problem where a set of $\Sc$ 
items (with cardinality $S$)
is given, each of them associated with a weight $\omega_{s}$ and a value $\nu_{s}$. 
The goal is to select a subset of items with maximum total value and total weight less or equal to 
a maximum given capacity, $\Omega$. 
Our reduction maps any given instance of the knapsack problem to a simpler, 
{\em special-case} instance of our own, as set forth next. 

The sets of L-nodes and I-nodes are, respectively, 
$\Lc = \{l_{1}\}$ and $\Ic = \{i_{1}\dots i_{S}\}$, 
i.e., there is only one L-node and as many I-nodes as  the number of items in the knapsack problem. 
Further, the L-node is connected with all the I-nodes.
We also set the number of epochs 
to an arbitrary number~$\hat{K}>0$, and the number of samples generated by each I-node to an arbitrary 
number~$r>0$.

Given the above, $\Pb$ is fixed and the decisions concern only  $\Qb$, which 
is now a vector with elements 
$q(i_s,l_1)$, mapping into the  $x_{s}$ variables in the  knapsack problem. Specifically, 
we activate  edge~$(i_s,l_1)$ in our problem if and only if $x_s=1$,  
i.e., $q(i_s,l_1) \gets x_s$. Furthermore, we map edge costs in our problem into 
item weights in the knapsack problem. In particular, let $\nu_{s}$ correspond to the opposite 
of the link cost $c_{i_s,l_1}$,
then we have  
a perfect correspondence between the objective 
of the knapsack problem and that in \Eq{obj}.

Next, we need to map the capacity constraint 
in the knapsack problem to constraint \Eq{constr}. 
To this end, we first set $T^{\max}\gets\infty$.  
Then, given that $\Pb$ is fixed, 
$\gamma=1$, and the L-node 
can only receive data from any number of I-nodes, the amount of data received by L-node $l_1$ 
in each epoch is
$r$ or $0$, depending on the value of $x_{s}$. 
A correspondence between the constraint in the knapsack problem  and that in our problem is 
then established by fixing $\epsilon^{\max} \gets \Omega$, and setting 
the $c_1$--$c_3$ coefficients in \Eq{generic-law} in such a way that setting~$x_s$ 
to~$1$ results in an increase of learning quality of~$\omega_s$, i.e.,
\begin{equation}
\nonumber
\frac{\log(c_3+X+X_s)-\log(c_3+X)}{\sqrt{K}}c_2=\omega_s,
\end{equation}
where $X_s=\frac{r(K+1)}{2}$~is the increase in the expected number of samples 
obtained by using I-node~$i_s$.
In other words, the equation above represents the increase in the 
value of learning quality (see \Eq{generic-law}) obtained by activating~$i_s$; 
we thus set the parameters of our problem so that the above increase results to be equal to 
the weight~$\omega_s$ assigned to~$s$ 
in the knapsack problem.

Last, we need the reduction to take (at most) polynomial time. In our case, it is straightforward 
to see that the mapping takes linear time, namely~$O(|\Lc|+|\Ic|)$, hence, the condition is fulfilled.

In summary, any instance of an NP-hard problem can be transformed into a special-case 
instance of our own, which proves the thesis.
\end{IEEEproof}

In spite of its complexity, the problem of minimizing \Eq{obj} subject to
constraint \Eq{constr}  presents several features that can be
exploited to solve it efficiently and effectively. 
Specifically, both the objective in \Eq{obj} and the constraint in
\Eq{constr} are submodular (intuitively, the set-wise equivalent of
convex~\cite{lovasz1983submodular}). Submodular optimization problems
can often be solved with polynomial- or even linear-time greedy
algorithms, with very good, even constant, competitive
ratios~\cite{conforti1984submodular}. 
Notice how both the results in~\cite{lovasz1983submodular} and~\cite{conforti1984submodular} are 
presented with reference to abstract, generic problems where the goal is to select some elements 
from  set~$\Xc$, with no reference (hence, no reliance) on specific scenarios.

Let us indicate with~$f(\Yc)$ the {\em objective function} in 
\Eq{obj}, and with~$g(\Yc)$ the {\em constraint} in \Eq{constr}. In
our case, the set~$\Xc$ of elements to choose from is given by
$\Xc=\Lc\times\Lc\cup\Lc\times\Ic$, i.e., the set of possible I-L and
L-L edges we can create, and~$\Yc$ is the subset 
of actually selected edges. 

The objective~$f(\Yc)$  and constraint~$g(\Yc)$ of our problem have several interesting and useful properties. Concerning the former, it is possible to prove the following result.
\begin{property}
\label{prope:obj-submodular}
The objective function in \Eq{obj} is submodular and non-decreasing.
\end{property}
\begin{IEEEproof}
Let~$j=(a,b)$ be an edge in our logical topology graph, with~$a\in\Lc$
and~$b\in\Lc\cup\Ic$; let~$\Sc\subset\Xc$ be the set of currently selected edges. By adding~$j$, we incur the per-edge
communication cost~$c_{a,b}$; also, we may incur per-node
operational costs~$c_a$ or $c_b$, depending on whether or not there
are already edges in~$\Sc$ with $a$ or~$b$ as endpoints.
Similar arguments hold for the cost of adding~$j$ to~$\Tc\supset
\Sc$. Thus, 
\begin{eqnarray}
\nonumber
f(\Sc\cup\{j\})-f(\Sc)&=&c_{a,b}+c_a\ind{a\not\in \Sc}+c_b\ind{b\not\in \Sc}\\
\nonumber
f(\Tc\cup\{j\})-f(\Tc)&=&c_{a,b}+c_a\ind{a\not\in \Tc}+c_b\ind{b\not\in \Tc}.
\end{eqnarray}
Since $\Sc$~is a subset of~$\Tc$, it also holds that~$\ind{a\not\in
  \Sc}\geq\ind{a\not\in \Tc}$ and~$\ind{b\not\in \Sc}\geq\ind{b\not\in
  \Tc}$, from which it follows that~$f(\Sc\cup\{j\})-f(\Sc)\geq
f(\Tc\cup\{j\})-f(\Tc)$, i.e., the very definition of submodularity \cite{lovasz1983submodular}.
The fact that \Eq{obj} is non-decreasing trivially comes from the observation that, as more I-L or L-L edges are added, the cost always increases. 
\end{IEEEproof}

As for the constraint, the analysis is a little more complex, and we
perform it separately for I-L and L-L edges,
proving first \Prope{edges-il} concerning the former, and then \Propo{edges-ll} concerning the latter.

For simplicity of notation, 
we drop the dependency on $\Pb$ and $\Qb$ while presenting our
derivations. 
%
\begin{figure}[th]
	\centering
	\centering
	\includegraphics[width=.7\columnwidth]{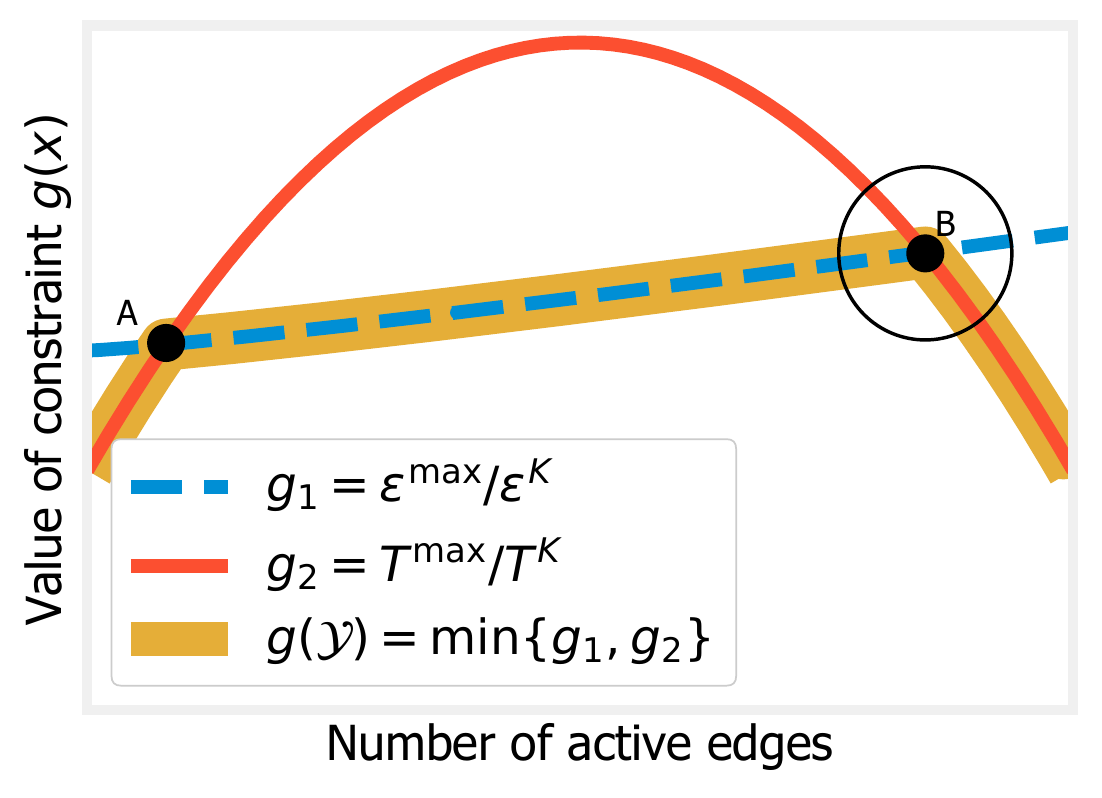}
	\caption{Qualitative example of the constraint in \Eq{constr} and its
		components.
		\label{fig:g-quali}
	} 
    \vspace{-3mm}
\end{figure}

\begin{property}
\label{prope:edges-il}
When the choices are limited to I-L edges, i.e.,~$\Xc=\Lc\times\Ic$,
then the constraint in \Eq{constr} is submodular and has exactly one maximum.
\end{property}
\begin{IEEEproof}
Let us study the two parts of the constraint \Eq{constr} separately,
writing~$g_1=\frac{\epsilon^{\max}}{\epsilon^K}$,
$g_2=\frac{T^{\max}}{T^K}$, and~$g(\Yc)=\min\{g_1,g_2\}$, as
exemplified in \Fig{g-quali}. 
From \Eq{generic-law}, $g_1=\frac{\epsilon^{\max} \sqrt{\gamma K}}{c_1 \sqrt{\gamma K}+c_2 \log{\left( c_3+X\right) }}$, and its second derivative is
$
\frac{c_2 \epsilon^{\max}  \sqrt{ \gamma K} }{{{\left( c_3+X\right) }^{2}} {{\left( c_1 \sqrt{\gamma K}+c_2 \log{\left( c_3+X\right) }\right) }^{2}}}{+}
\frac{2 {{c_2}^{2}} \epsilon^{\max} \sqrt{ \gamma K}}{{{\left( c_3+X\right) }^{2}} {{\left( c_1 \sqrt{\gamma k}{+}c_2 \log{\left( c_3+X\right) }\right) }^{3}}}.
$ 
Such a derivative is always positive, hence, $g_1$ is submodular.

The behavior of~$g_2$ is more complex: we know from 
\Eq{generic-law} 
that the number of epochs decreases as $X$ increases, according
to an inverse-log law. 
Also,   as shown in \Sec{characterize}, $\tau_l^k(t)$
and $\dv{H^k(t)}{t}$  are
proportional to $X_l^k$  and $\prod_{l\in\Ic} X_l^k$, respectively. 
Thus, $T^K$ is proportional to $K$ and $\prod_{l\in\Ic} X_l^k$. 
Replacing $K$ with \Eq{generic-law}, we get that $T^K$ 
behaves like~$\frac{\prod_{l\in\Ic}  X^k_l}{\log X}$, i.e., it can be shown that it
decreases until it reaches a minimum, and then increases. 
It follows that $g_2=\frac{T^{\max}}{T^K}$ is concave, hence, submodular.

Looking now at~$g(\Yc)$, the minimum of two submodular functions is
not guaranteed to be submodular in general; however, since~$g_1$ is
not only submodular but also monotonically increasing, the
submodularity of~$g_2$ also implies
that~$g(\Yc)$ as a whole is submodular~\cite{lovasz1983submodular}. 
Next, consider the maximum of~$g(\Yc)$, with the latter being  equal to
$\min\{g_1,g_2\}$. As exemplified in \Fig{g-quali}, we know that~$g_1$
starts from a value close to~$\epsilon^{\max}$ and then monotonically
increases towards infinity, while~$g_2$ starts with a small value,
increases until it has a global maximum, and then decreases again. If
$g_2$ is always smaller than~$g_1$, then~$g(\Yc)=g_2$ has exactly one
global maximum, consistently with the hypothesis. If they cross (as
in \Fig{g-quali})), they do so in exactly two points, say 
$A$ and $B$, such that the maximum of~$g_2$ is between $A$ and $B$. Then, the following holds:
(i) before $A$, $g(\Yc)=g_2$, which is increasing before its maximum;
(ii) between $A$ and $B$, $g(\Yc)=g_1$, which is always increasing;
 (iii) after $B$, $g(\Yc)=g_2$ and, since we are after its maximum, $g(\Yc)$ is decreasing -- hence, $B$ is $g(\Yc)$'s only maximum.
Therefore, in all cases $g(\Yc)$ is submodular and has exactly one
maximum, and, until such a maximum is reached, $g(\Yc)$ is also monotonically non-decreasing.
\end{IEEEproof}

As for L-L edges,
their influence on the learning process can be quantified by studying
the graph they form.
Specifically, we are able to state the following result concerning regular graphs:
\begin{proposition}
\label{propo:edges-ll}
When the choices are limited to sets of L-L edges such that the graph created by L-nodes is uniform, then the constraint \Eq{constr} is submodular and has exactly one maximum.
\end{proposition}
\begin{IEEEproof}
The arguments in support of \Propo{edges-ll} can be summarized as follows:
1) the error reached after a given number~$K$ of epoch is proportional to~$1/\gamma$~\cite[Eq.~(7)]{neglia};
2) the learning time  is proportional to~$1/\gamma$~\cite[Eq.~(18)]{neglia};
3) for random regular graphs, the relationship between the spectral gap and the graph degree has been shown~\cite{vu2008random,tikhomirov2019spectral} to follow a square-root law, which is concave.
Recalling that concavity is the continuous equivalent of submodularity, the first part of the proposition follows.
The second part follows from the fact that \Eq{constr} is the minimum between a monotonic function
(as we add more L-L edges, the error decreases) and a function with at most one maximum (the inverse of the learning time, which decreases until an optimal degree is reached and then increases, as shown in \cite{neglia}).
\end{IEEEproof}

Combining \Prope{edges-il} and \Propo{edges-ll}, we can now prove the following result: 
\begin{corollary}
When the graph created by L-nodes is uniform, constraint \Eq{constr} is submodular and has exactly one maximum.
\end{corollary}
\begin{IEEEproof}
The possible actions are either adding an I-L edge, or an L-L one. As per (respectively) \Prope{edges-il} and \Propo{edges-ll}, both actions preserve the submodularity property, and result in a function with exactly one maximum.
\end{IEEEproof}

\section{The DoubleClimb Algorithm}
\label{sec:algo}


We now seek to solve the problem stated in \Sec{problem}, i.e.,
determining the~$\Pb$, $\Qb$ and~$K$ resulting in the lowest cost
\Eq{obj} subject to the constraint in \Eq{constr}, in a practical and
efficient way. To this end, we first extend existing results on the
performance of greedy algorithms when optimizing submodular problems,
in \Sec{sub-solving}. Based on such results, we present our own
DoubleClimb algorithm in \Sec{sub-doubleclimb}, and analyze its
properties in \Sec{dc-analysis}.

\begin{algorithm}[b]
\caption{Greedy algorithm for submodular problems\label{alg:greedy}}
\begin{algorithmic}[1]
\State{$\Sc\gets\emptyset$}
\While{$g(\Sc)\geq c$}
 \State{$j^*\gets\arg\min_{\Xc\setminus\Sc}\frac{c_j}{g(\Sc\cup\{j\})-g(\Sc)}$} \label{line:greedy-argmax}
 \State{$\Sc\gets\Sc\cup\{j\}$} \label{line:greedy-add}
\EndWhile
\State\Return{$\Sc$}
\end{algorithmic}
\end{algorithm}

\subsection{Greedy solutions to submodular problems} 
\label{sec:sub-solving}

Let us consider 
\Alg{greedy}, 
which solves submodular problems with non-decreasing objective and constraints. More
formally, it 
selects a subset~$\Sc\subseteq\Xc$ of elements subject to
a submodular non-decreasing
constraint~$g(\Sc)\geq 1$, while minimizing a
submodular non-decreasing cost function~$f(\Sc)$. At every iteration, \Line{greedy-argmax} selects
the element minimizing the cost to benefit ratio 
$\frac{f(\Sc\cup\{j\})-f(\Sc)}{g(\Sc\cup\{j\})-g(\Sc)}$; such an element is then added to $\Sc$ (\Line{greedy-add}).
As shown in~\cite[Thm.~4.7]{iyer2013submodular}, \Alg{greedy} is
$1+\frac{1}{|\Xc|}$-competitive.
However, the original proof requires both the objective and the
constraint to be submodular and non-decreasing. In our case,
\Prope{edges-il} and \Propo{edges-ll} prove {\em weaker} properties, in that our constraint is not 
guaranteed to be non decreasing, as in \Fig{g-quali}; 
therefore, the result in~\cite{iyer2013submodular} cannot immediately be applied to our problem.

None the less, it is possible to prove that a less restrictive condition than being non-decreasing, namely, having only one maximum, is sufficient for the result to hold:
\begin{property}
\label{prope:enough}
If $f(\Yc)$~is submodular non-decreasing and~$g(\Yc)$ is submodular and has
only  one maximum, then the above algorithm minimizes~$f(\Yc)$ s.t.~$g(\Yc)>0$, with a competitive ratio of~$1+\frac{1}{|\Xc|}$.
\end{property}
\begin{IEEEproof}
The property generalizes the results
in~\cite[Thm.~4.7]{iyer2013submodular}. 
The proof therein follows from analyzing the steps of the above algorithm until its
convergence, and leveraging the fact that the sequences of marginal
cost increases and constraint improvements are (resp.)
monotonically non-decreasing and monotonically non-increasing. 
This is of course true if, as in the original hypotheses, $g(\Yc)$~is
monotonically non-decreasing. However, this also holds if $g(\Yc)$ 
has only one maximum, as per the hypothesis of our property. This is
because, if the algorithm 
cannot find a feasible solution before the maximum of~$g(\Yc)$, i.e., as constraints become 
{\em closer} to being satisfied, it will also be impossible to find a feasible solution after the maximum, i.e., 
when constraints will get {\em farther} from being met. 
Thus, the sequences of marginal costs and improvements {\em of the selected elements} of~$\Xc$ have the required behavior. Indeed, the behavior of~$g(\Yc)$ for the non-selected items of~$\Xc$ {\em has no impact} on the validity of~\cite[Thm.~4.7]{iyer2013submodular}, nor of this property.
\end{IEEEproof}

\subsection{Algorithm description}
\label{sec:sub-doubleclimb}

\Prope{enough} implies that the algorithm in \Sec{sub-solving} could efficiently select the L-L links~$\Pb$
and the I-L links~$\Qb$, i.e., which L-L nodes cooperate with one another and which information they can leverage
{\em if such decisions could be made independently}, without one impacting the other.
  However, they are clearly interlinked;
thus, we propose a more complex solution strategy, called
DoubleClimb, which operates as follows. 
\begin{itemize}
    \item First, based on the nodes 
capabilities
      defined in \Sec{model}, 
      DoubleClimb  determines  $\Pb$ and $\Qb$. It does so by selecting I-L and L-L edges in two nested
      loops, with L-L edges resulting in a uniform graph \cite{neglia}. It also 
 selects the most appropriate value of~$K$ for each set of selected edges.

\item Given such decisions, it computes the
  system performance characterized in \Sec{characterize}, thus yielding 
the   error~$\epsilon^K(\Pb,\Qb)$, the learning 
  time~$T^K(\Pb,\Qb)$, as well as the cost~$C^K(\Pb,\Qb)$.

    \item It then compares the obtained values for the learning
      time and error  against the limits~$\epsilon^{\max}$ and $T^{\max}$, and
      evaluates whether a sufficiently low cost has been achieved. If
      so,  DoubleClimb returns the problem solution; otherwise, it tries to
      improve the decisions until the system constraints  are
      met and the cost  is further reduced. 
\end{itemize}
Intuitively, we begin with a sparsely connected graph with no L-L and no I-L edges, and then we keep increasing the connectivity until the constraints are satisfied.

\begin{algorithm}[th]
\caption{The DoubleClimb algorithm
\label{alg:doubleclimb}
} 
\begin{algorithmic}[1]
\State{$d_{\text L}\gets 0$} \label{line:init-l}
\State{$\texttt{best\_sol}\gets\emptyset$} \label{line:init-bs}
\While{$d_{\text L}<|\Lc|$}
 \State{$d_{\text L}\gets d_{\text L}+1$} \label{line:increment-l}
 \State{$\texttt{ll}\gets\textbf{cheapest\_uniform}(d_{\text L})$} \label{line:choose-ll}
 \State{$\texttt{il}\gets\emptyset$} 
 \While{\Eq{constr} is not verified $\wedge\texttt{il}\not\equiv\Ic\times\Lc$} \label{line:more-il}
  \State{$i^*,j^*{\gets}\arg\min_{i,l}\frac{c_{i,l}}{g(\texttt{il})-g(\texttt{il}\cup\{(i,l)\})}$} \label{line:choose-il}
  \State{$\texttt{il}\gets\texttt{il}\cup\{(i^*,j^*)\}$} \label{line:add-il}
 \EndWhile
 \If{$C^\text{curr}< C^\text{best}$} \label{line:check-ok}
  \State{$\texttt{best\_sol}\gets\texttt{ll}\cup\texttt{il}$} \label{line:update-bs}
 \ElsIf{$C^\text{curr}_\text{LL}>C^\text{best}_\text{LL}\wedge C^\text{curr}_\text{IL}> C^\text{best}_\text{IL}$} \label{line:check-stop}
  \State{\textbf{break}} \label{line:break}
 \EndIf
\EndWhile
\State\Return{\texttt{best\_sol}} \label{line:return}
\end{algorithmic}
\end{algorithm} 

The DoubleClimb algorithm is presented in
\Alg{doubleclimb} and detailed below. 
It begins (\Line{init-l}) by setting to zero the degree $d_{\text L}$ of the subgraph made of L-L edges, and to the empty set the best solution \path{best_sol}. Then, while~$d_{\text L}<|\Lc|$, i.e., while such a subgraph is not a clique, $d_{\text L}$ is first incremented by one (\Line{increment-l}), and then the cheapest L-L uniform subgraph of degree~$d_{\text L}$ is chosen in \Line{choose-ll}.

Given such a choice of L-L edges, the algorithm selects the I-L edges
 essentially in the same way as described in \Sec{sub-solving}: for all possible edges, the cost/benefit ratio -- i.e., the ratio between the cost of adding the edge and how closer to feasibility the problem becomes by doing so -- is computed in \Line{choose-il}, and the edge associated with the lowest ratio is chosen. The loop continues until either all I-L edges are exhausted, or a feasible solution, satisfying constraint \Eq{constr}, is found (\Line{more-il}).
In the latter case, the cost of the current solution $C^\text{curr}$, 
computed as per \Eq{cost}, 
is compared to the one of the best solution found so far ($C^\text{best}$); note that, by convention,
the cost of the empty set is equal to $\infty$.  If warranted, the
best solution is updated (\Line{update-bs}), otherwise  
we perform the check in \Line{check-stop} to assess whether other
solutions should be explored. 
Indeed, as proven in \Propo{noloss} below, the
submodularity of costs implies that trying higher values of $d_{\text L}$ does not lead to cheaper solutions.

If neither happens, the next value of $d_{\text L}$ is tried. After all
values of $d_{\text L}$ are exhausted,
the best solution \path{best_sol} is returned in \Line{return}. If no feasible solution has been found, the problem instance is infeasible and the algorithm returns~$\emptyset$.

\subsection{Algorithm analysis}
\label{sec:dc-analysis}

We now prove that \Alg{doubleclimb} has an excellent competitive ratio
as well as low complexity. 
As first step, we show that the stopping condition in \Line{check-stop} is valid, i.e., no solution better than \path{best_sol} is ignored by halting the algorithm when the condition is met.
\begin{proposition}
\label{propo:noloss}
If the condition specified in \Line{check-stop} of \Alg{doubleclimb} is met, then no solution cheaper than \path{best_sol} will be found for higher values of $d_{\text L}$.
\end{proposition}
\begin{IEEEproof} 
Let~$d_{\text L}^\text{best}$ be the value of $d_{\text L}$ for which the current
best solution was found, and $C^\text{best}_\text{LL}$ and
$C^\text{best}_\text{IL}$ the corresponding costs for L-L and I-L
edges (resp.). At the current iteration, we have
$d_{\text L}=L^\text{curr}>L^\text{best}$, and the corresponding costs are
$C^\text{curr}_\text{LL}>C^\text{best}_\text{LL}$ and
$C^\text{curr}_\text{IL}>C^\text{best}_\text{IL}$. 
Let us now consider a future iteration where the value of $d_{\text L}$ is
$d_{\text L}^\text{next}>d_{\text L}^\text{curr}>d_{\text L}^\text{best}$. 
$C^\text{next}_\text{LL}$ depends on two effects: if we increase the
number of L-L edges, the  cost
due to L-L edges will increase. However, more L-L edges also 
imply fewer epochs, thus they may lead to a reduced cost.
Since similar observations hold for $C^\text{next}_\text{IL}$,  
which effect prevails depends on how strong the benefit of increasing
$d_{\text L}$ is. 
However, as per the submodularity property (\Propo{edges-ll}), the
benefit of adding L-L edges and I-L edges decreases as $d_{\text L}$
increases: if moving from $d_{\text L}^\text{best}$ to $d_{\text L}^\text{curr}$
actually increased the cost of L-L and I-L edges, it is not possible
that moving to $d_{\text L}^\text{next}$ will provide a better solution.
\end{IEEEproof}

Thanks to \Propo{noloss} and \Prope{enough}, we can now prove our main result 
about the {\em competitive ratio} of \Alg{doubleclimb}, i.e., the ratio of the cost of the solution it yields to the one of the optimal solution.
\begin{theorem}
\Alg{doubleclimb} has $1+\frac{1}{|\Ic|}$ competitive ratio.
\end{theorem}
\begin{IEEEproof}
There are two possible sources of suboptimality, namely, the choice of
$d_{\text L}$ and that of the I-L edges to select. By \Propo{noloss} and
considering that, if the condition in \Line{check-stop} is never
triggered,  all possible values of $d_{\text L}$ are tried out, the
choice of $d_{\text L}$ is optimal. 
As for the I-L edges, \Line{more-il}--\Line{add-il} of
\Alg{doubleclimb} reflect exactly the same algorithmic steps reported
in \Sec{sub-solving} which, as per \Prope{enough}, lead to a $1+\frac{1}{|\Ic|}$ competitive ratio in our case.
\end{IEEEproof}

Finally, we can prove that \Alg{doubleclimb} has a very low, namely,
cubic {\em worst-case} computational complexity. 
\begin{property}
\label{prope:complexity}
\Alg{doubleclimb} has a worst-case computational complexity of $O(|\Lc|^2|\Ic|)$.
\end{property}
\begin{IEEEproof}
From inspection of the nested loops in \Alg{doubleclimb}, one can
see that the outer one is run at most once for each value of $d_{\text L}$, i.e., at most $|\Lc|$~times. The inner one is ran at most once for each possible I-L edge, i.e., at most~$|\Lc||\Ic|$ times.
As for the set of edges to activate for each value of~$d_L$ (function~$\textbf{cheapest\_uniform}$ in \Line{choose-ll}), they can be pre-computed and thus do not influence the overall complexity. 
\end{IEEEproof}
It is also worth  stressing that \Prope{complexity} concerns the {\em
  worst-case} complexity, but the actual one is often much lower. Indeed, in \Line{more-il}--\Line{add-il} we are likely to compute the same costs in different iterations; if such costs are cached, {\em \`{a} la} dynamic programming, run time can be dramatically decreased, to be slightly more than linear in~$|\Lc|+|\Ic|$.

\section{Numerical Results}
\label{sec:results}

In the following, we describe the reference scenario and benchmark solutions we consider (\Sec{refscen}), before studying the performance of DoubleClimb (\Sec{resultsresults}).

\subsection{Reference scenario}
\label{sec:refscen}

We consider an Internet-of-things (IoT) environment similar to the one referred in~\cite{wang2019adaptive}, whereby:
\begin{itemize}
    \item individual sensors produce samples, either periodically or as a reaction to an external event;
    \item {\em aggregators}, also known as gateways, collect and summarize the samples, before forwarding them in uplink;
    \item distributed ML algorithms, running at the edge of the network, leverage the samples to gather insights on 
    the changes in external conditions.
\end{itemize}
In terms of our system model, aggregators correspond to I-nodes, and
edge nodes running the ML algorithms correspond to L-nodes. 
New samples arrive every few seconds, and updating the gradient
computations takes a comparable time. 
Note that similar approaches have been proposed for such applications as
smart-city monitoring~\cite{valerio2018energy}, support of connected
vehicles~\cite{ye2018machine}, and attack/anomaly
detection~\cite{diro2018distributed}.  

With reference to the taxonomy
in \Sec{intro}, we fall in the {\em active learning} case, as the data
arrival and gradient computation are interleaved but not synchronized,
e.g., new data can arrive both before and after a gradient computation
is complete. 

\begin{figure}[th]
	\centering
	\includegraphics[width=.7\columnwidth]{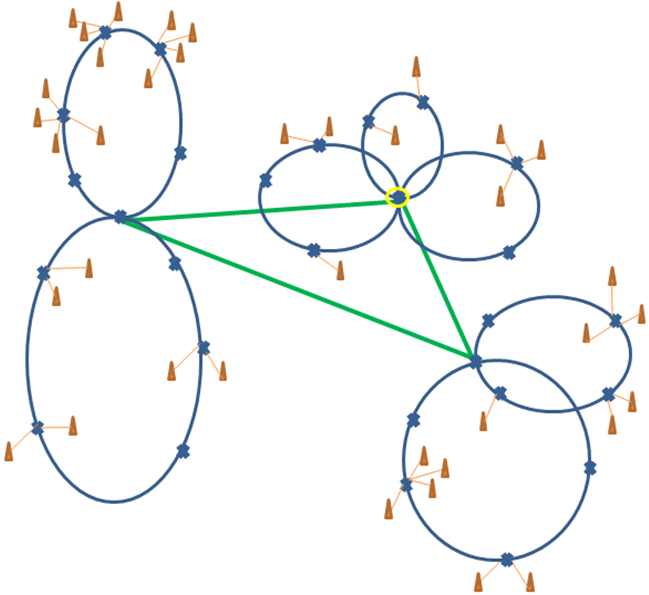}
	\caption{Our reference topology, depicting the network of a major operator (source: \cite{CrosshaulD12}).
	} 
	\label{fig:crosshaulTopology}
\end{figure}

We refer to the real-world urban topology presented 
in~\cite{CrosshaulD12} and shown in \Fig{crosshaulTopology},
depicting the network of a major  operator. Specifically, the
network nodes represented in brown act as
aggregators, hence, as I-nodes, while 
 those in blue are edge nodes acting as L-nodes.
As shown in \Fig{crosshaulTopology}, all L-nodes can be connected with one another, while each I-node can only be connected to one L-node.

Normalized sample  generation and gradient
computation times are distributed exponentially with mean~1, while the
I-L and L-L edges are randomly assigned a normalized  cost between 0 and 1 units. I- and L-nodes have 
  no operational cost, reflecting the fact that, in our reference
  environment, they cannot be switched off without discontinuing the
  service. 
In the  {\em basic} version of the scenario, at every epoch each I-node generates between 10 and 100 samples; such a value is proportional to the traffic served by each node in the real-world topology~\cite{CrosshaulD12}. In the {\em rich} scenario, representing applications where data is plentiful, such a value is multiplied by five.

{\bf Benchmark solutions.}
We compare DoubleClimb against two benchmark solutions. The first,
called Opt-Unif, follows the approach used (among others)
by~\cite{neglia}, and returns the cheapest solution among the feasible
ones such that both the graphs formed by L-L and I-L edges have uniform degree.

The second benchmark, labeled as ``Optimum/GA'' in the plots, performs the selection of the I-L edges 
(i.e., the inner loop in \Alg{doubleclimb}) leveraging a genetic algorithm (GA) 
approach with the following parameters:
\begin{itemize}
    \item number of generations: 50;
    \item solutions per population 100;
    \item parents mating: 4;
    \item mutation probability: 15\%;
    \item crossover type: single point;
    \item gene space: $\{0,1\}$;
    \item number of genes: $|\Ic||\Lc|$.
\end{itemize}
Each solution corresponds to a string of binary values whose length equals the number of 
possible I-L edges: having a $1$ in a given position means that the corresponding 
I-L edge is activated. The relatively large mutation probability reflects the importance of 
exploring multiple different solutions (i.e., exploration), given the combinatorial nature of 
the problem at hand and the fact that similar strings do not necessarily yield similar performance. 
When the size of the problem made it possible (i.e., $d_L\leq 6$), we have compared 
the performance of the genetic algorithm against the optimum obtained through brute force, 
and found that the two closely match.

\subsection{Learning tasks}
\label{sec:resultslearning}

We conduct our performance evaluation with reference to the two most relevant supervised 
learning tasks, namely:
\begin{itemize}
    \item a {\em classification} task on the famous MNIST digit database~\cite{digits};
    \item a {\em regression} task on the dataset used for the ITU AI~Challenge~\cite{ituchallenge}, 
    with the goal of predicting the throughput of a set of Wi-Fi nodes leveraging  their position 
    and settings.
\end{itemize}
Through these two datasets, we can show how our methodology works for the two most common and 
relevant types of supervised learning. 
We tackle both learning tasks via the virtually-ubiquitous tool of deep neural networks (DNNs). 
Specifically, we employ a fully-connected DNN including thee hidden layers, whose sizes  are 
100, 50, 20~neurons, respectively. The DNN is trained via stochastic gradient descent (SGD), 
with a learning rate of~$0.01$. All experiments are implemented in Python using 
the \texttt{pytorch} library.

As per the methodology presented in \Sec{sub-accuracy} and validated, among others, 
in~\cite{hestness2017deep}, we obtain the following values for
the parameters in \Eq{generic-law}:
\begin{itemize}
    \item for the classification task: $c_1=0.6799$, $c_2=0.4978$, $c_3=542.1$;
    \item for the regression task: $c_1=0.0956$, $c_2=0.5203$, 
    $c_3=963.2$.
\end{itemize} 
We quantify the goodness of fit through the mean square error (MSE) metric; 
in our experiments, the MSE for the classification and regression tasks is, 
respectively,~$0.0027$ and~$9.87\cdot 10^{-6}$.
It is interesting to remark that, while the $c_1$--$c_3$ parameters are quite different 
in the two cases, the error is remarkably small in both instances. 

In both cases, we profiled federated learning (FL) with~$|\Lc|=10$ learning nodes 
assisted by a central learning server, and varied the number of samples between~50\% and~100\% of 
the whole dataset. Local models are therefore averaged at every epoch, 
following the FedAvg~\cite{konen2015federatedOptimization} averaging strategy. 
Results have been averaged over 10~runs, changing the composition of the local datasets across different runs.

\subsection{Performance comparison}
\label{sec:resultsresults}

\begin{figure*}
	\centering
	\includegraphics[width=.24\textwidth]{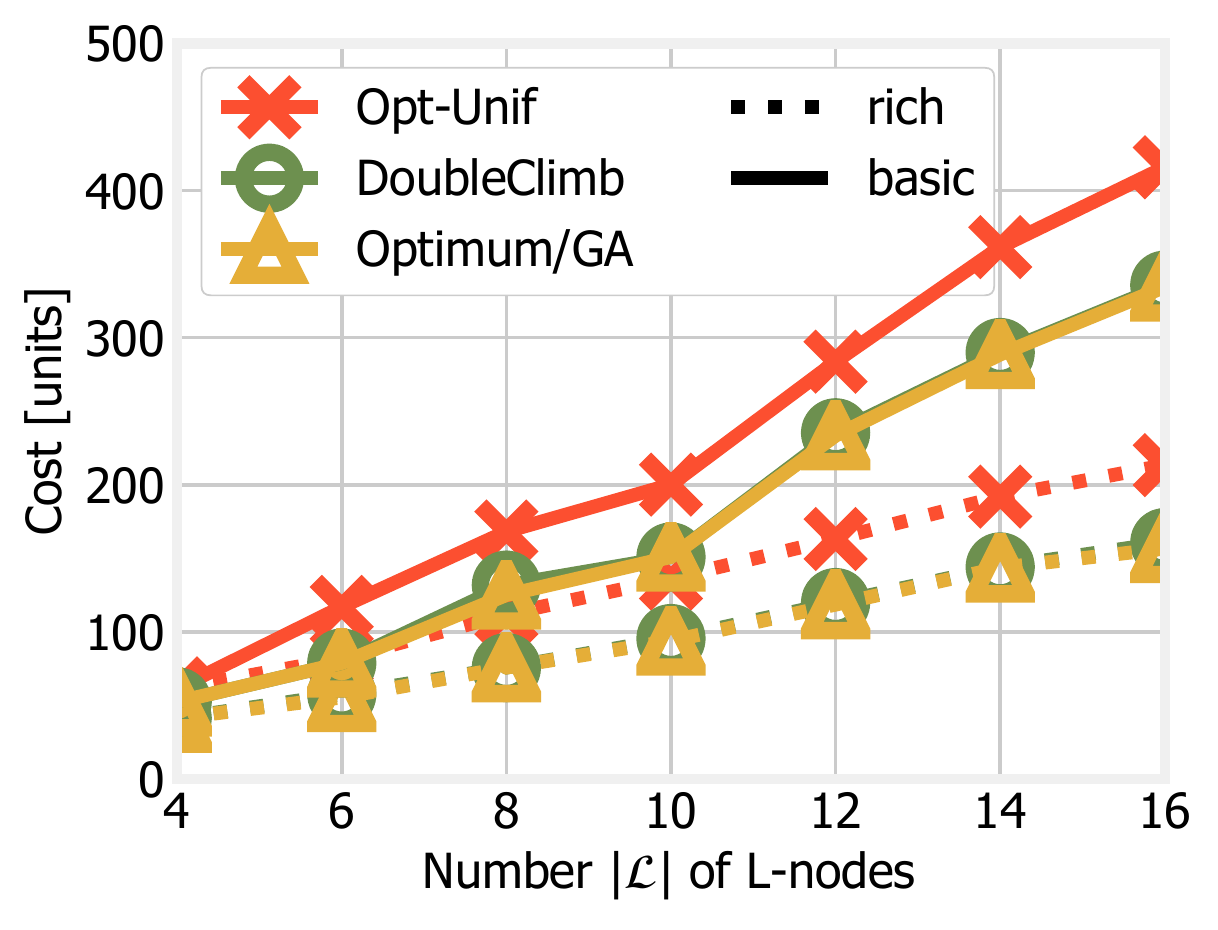}
	\includegraphics[width=.24\textwidth]{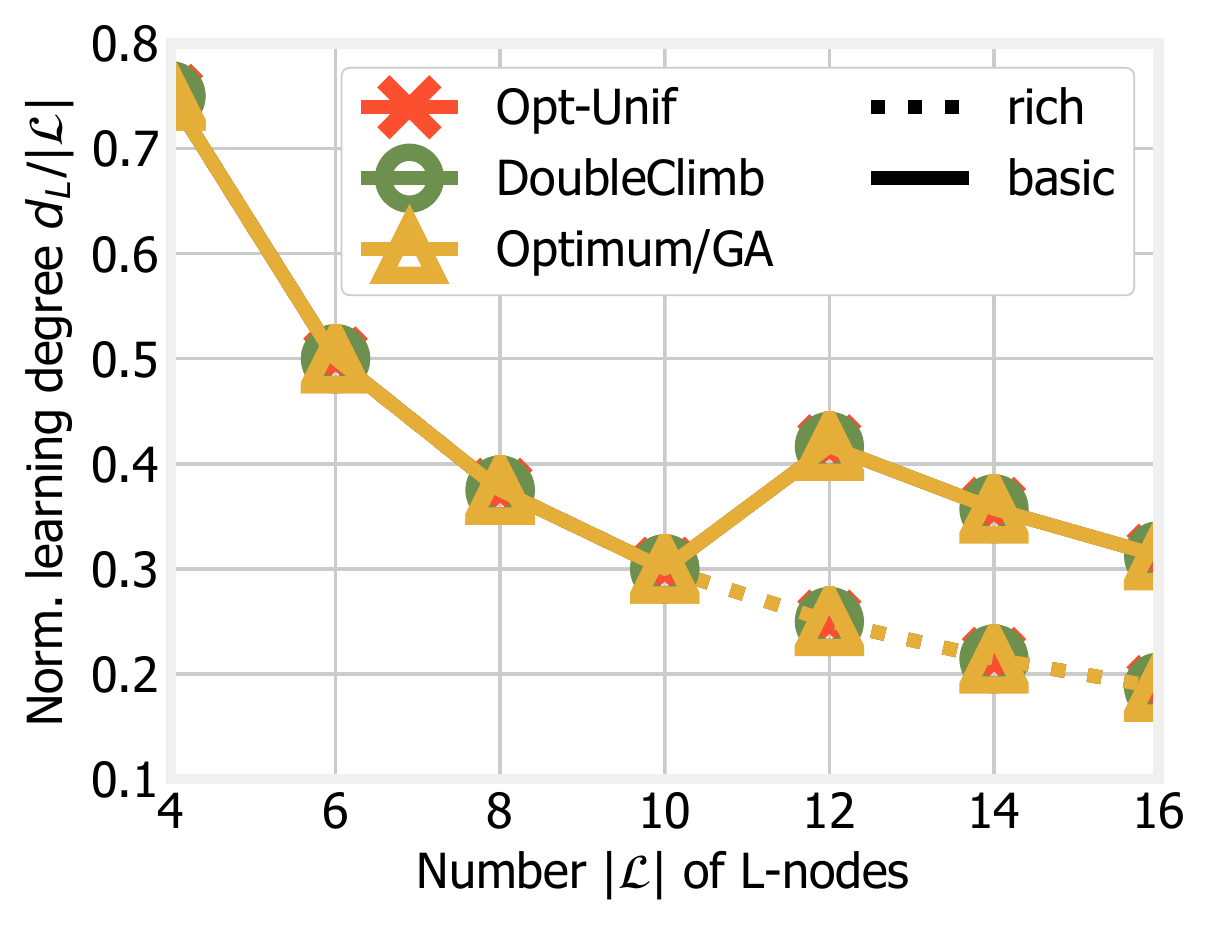}
	\includegraphics[width=.24\textwidth]{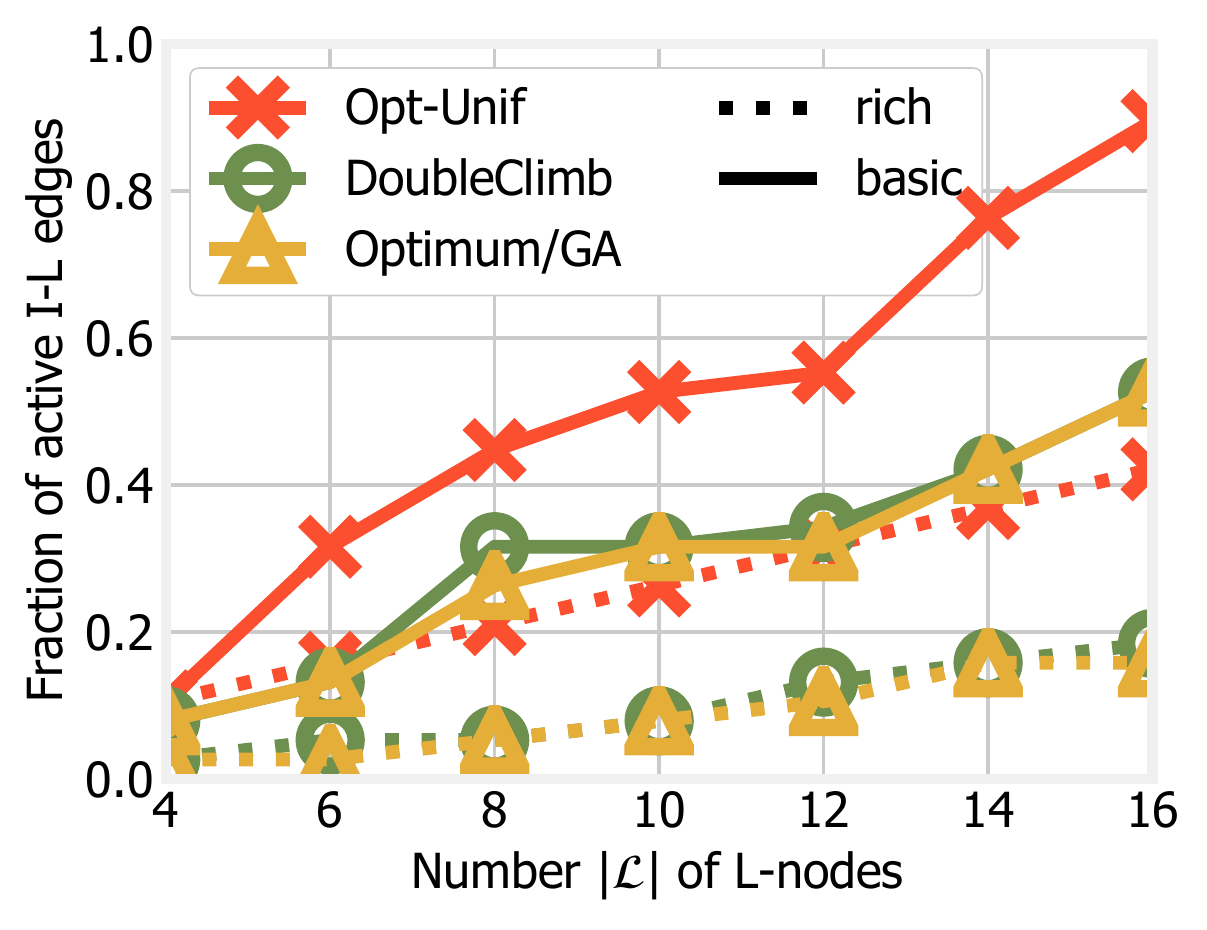}
	\includegraphics[width=.24\textwidth]{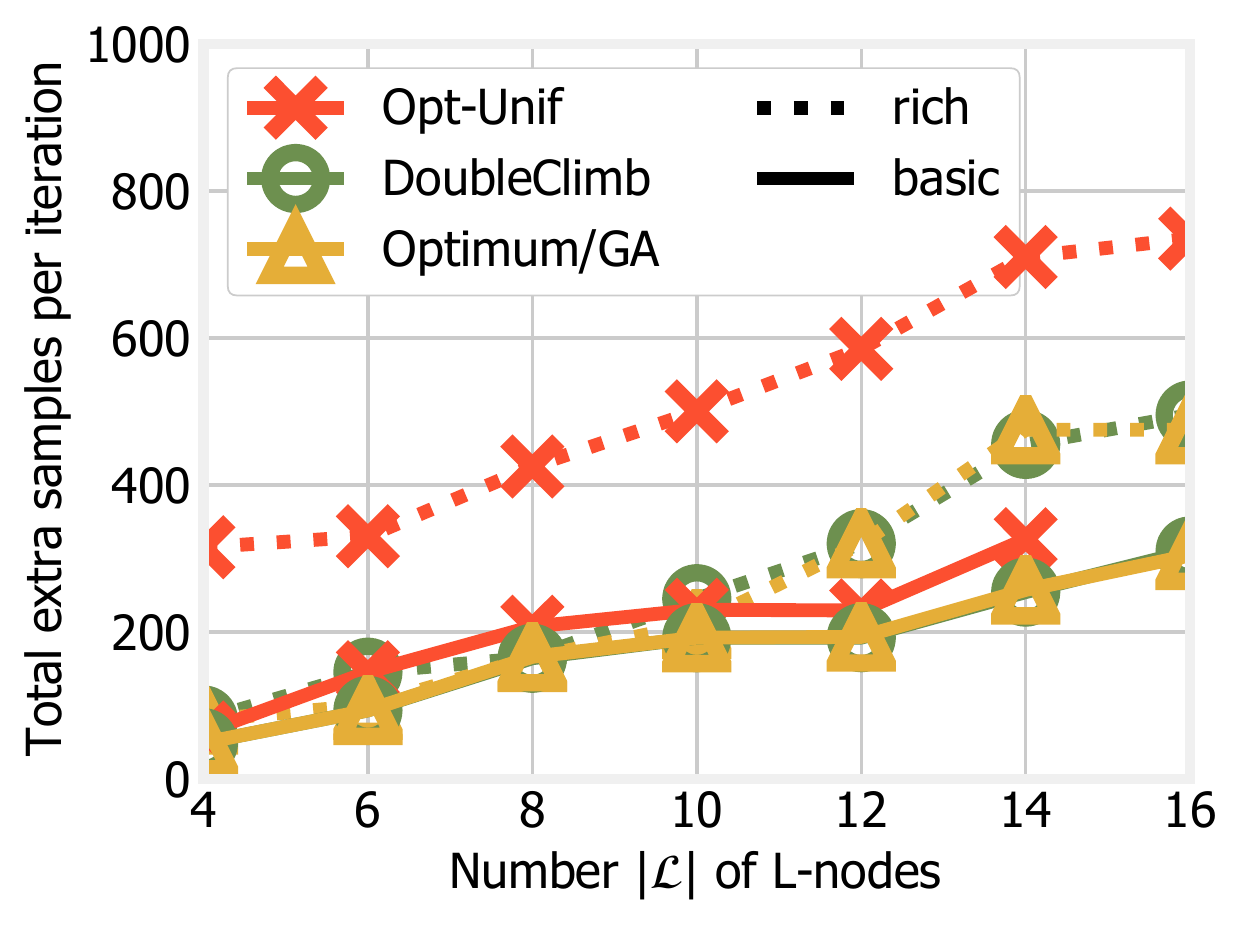}
	\caption{
		{\bf Classification task}: comparison between DoubleClimb, Opt-Unif and the optimum (obtained
		via brute-force) in the basic and rich scenarios, for different
		values of~$|\Lc|$. From left to right: total cost; selected value
		of $d_{\text L}$, normalized (to the
		maximum); fraction of selected I-L edges; total number of extra samples per epoch.
		\label{fig:compare}
	}
	\includegraphics[width=.24\textwidth]{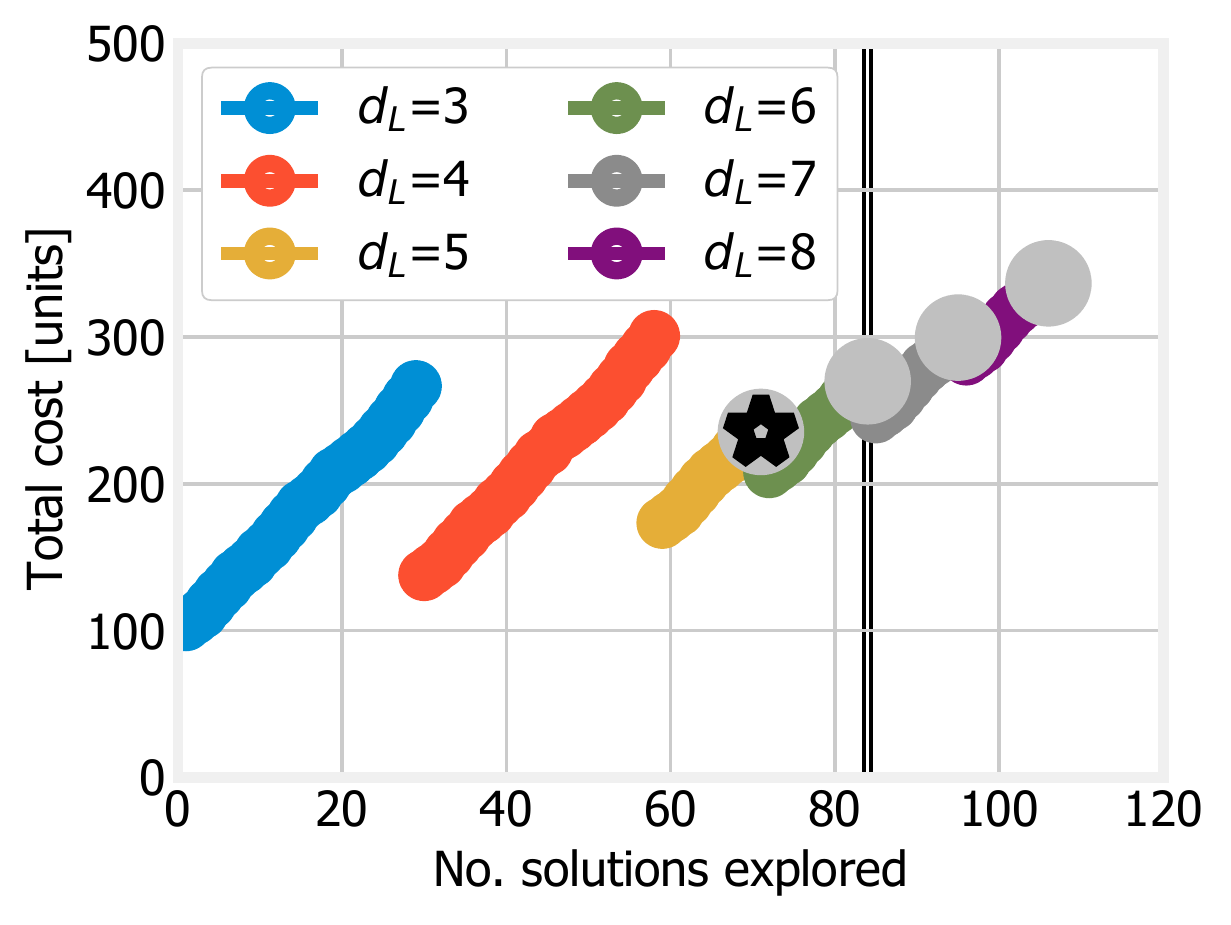}
	\includegraphics[width=.24\textwidth]{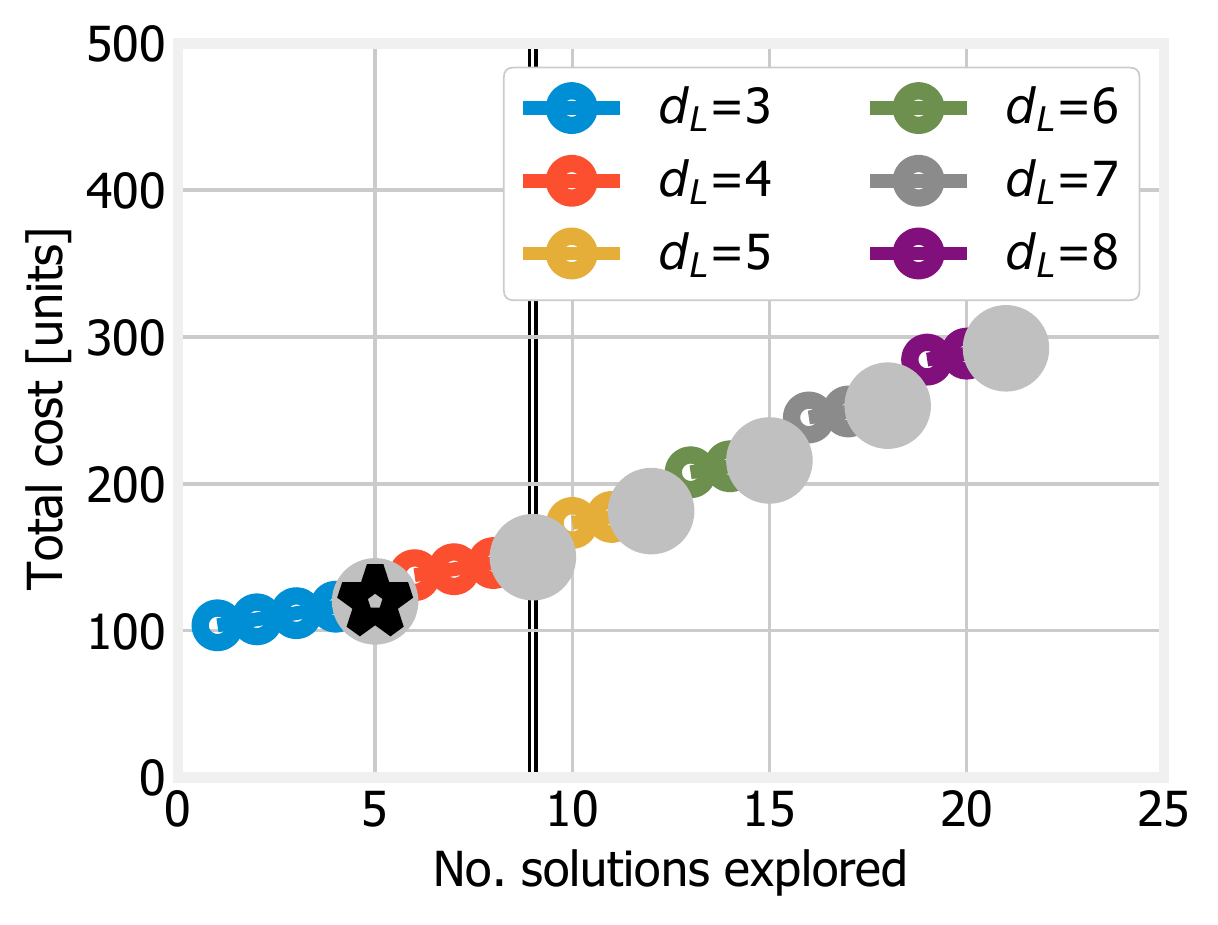}
	\includegraphics[width=.24\textwidth]{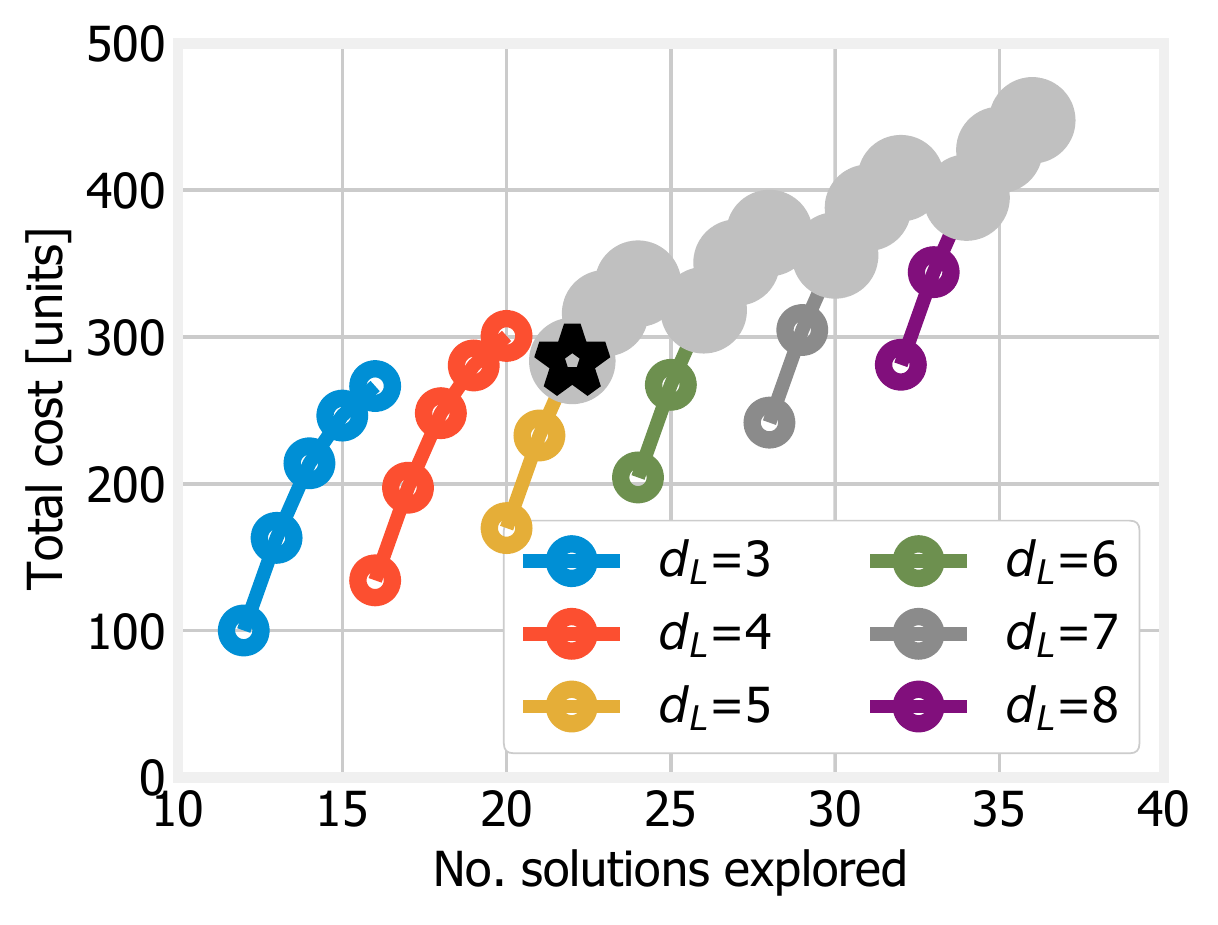}
	\includegraphics[width=.24\textwidth]{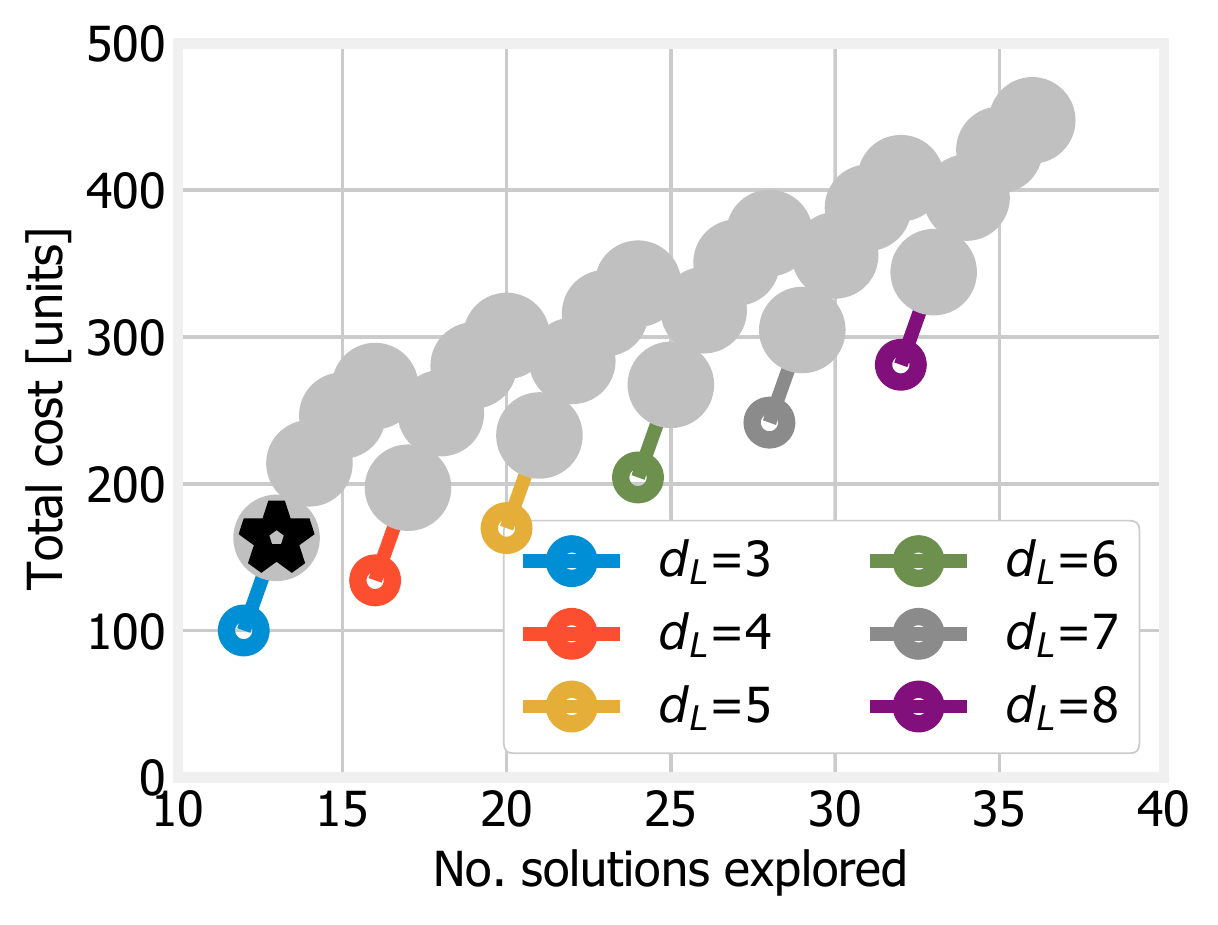}
	\caption{
		{\bf Classification task}: cost of the solutions examined at each iteration by DoubleClimb (first two plots) and Opt-Unif (last two plots), in the basic (first and third plot) and rich (second and fourth plot) scenarios.
		\label{fig:magicsticks}
	}
	\includegraphics[width=.32\textwidth]{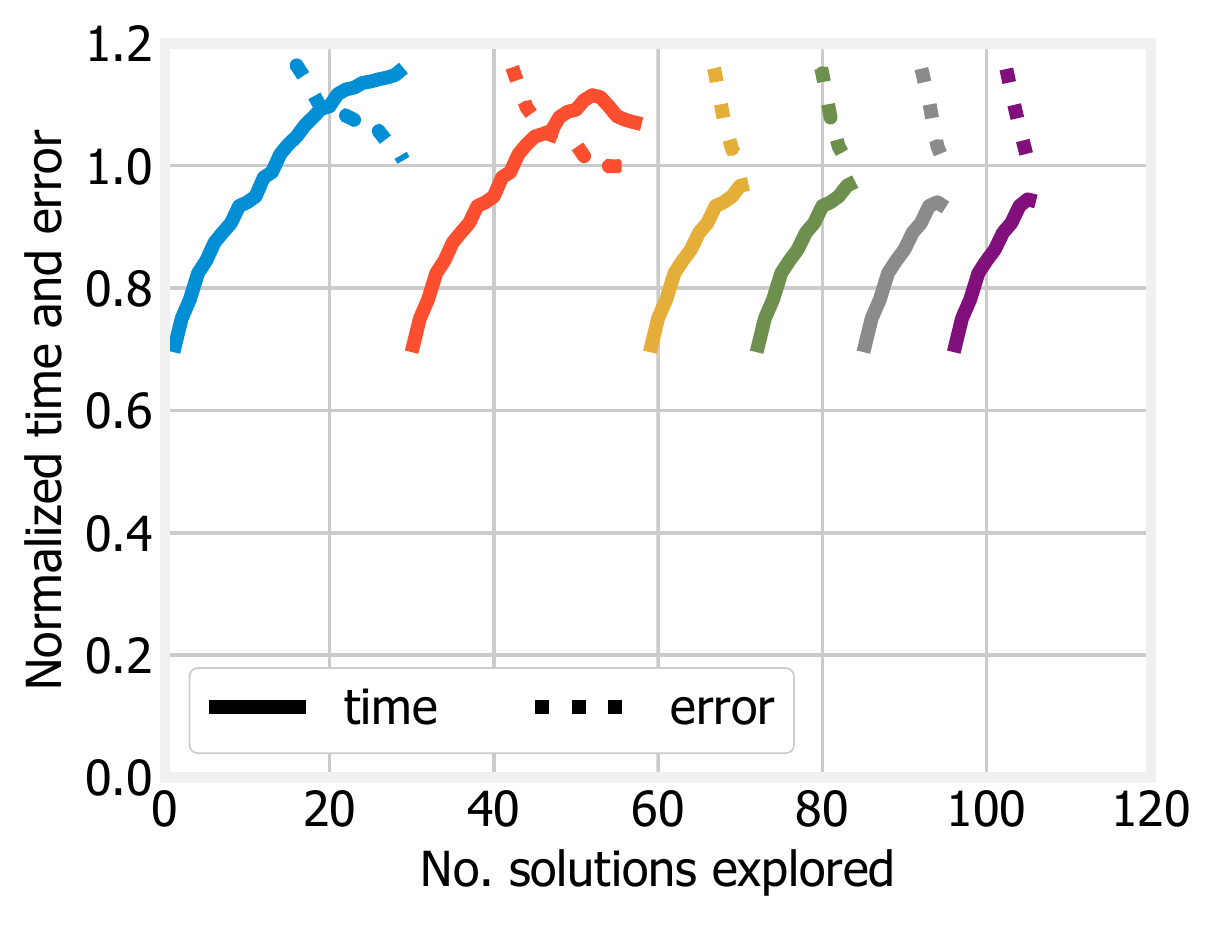}
	\includegraphics[width=.32\textwidth]{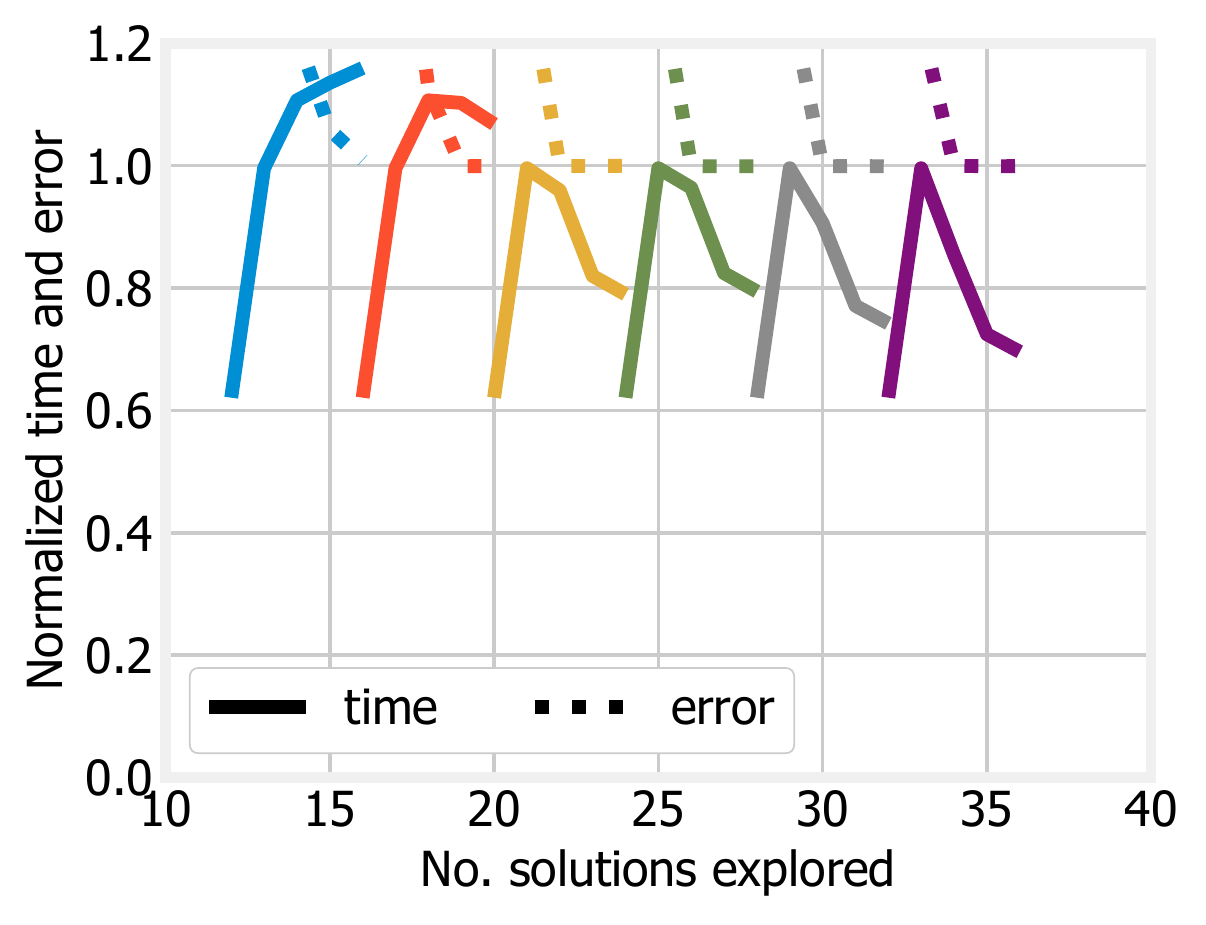}
	\includegraphics[width=.32\textwidth]{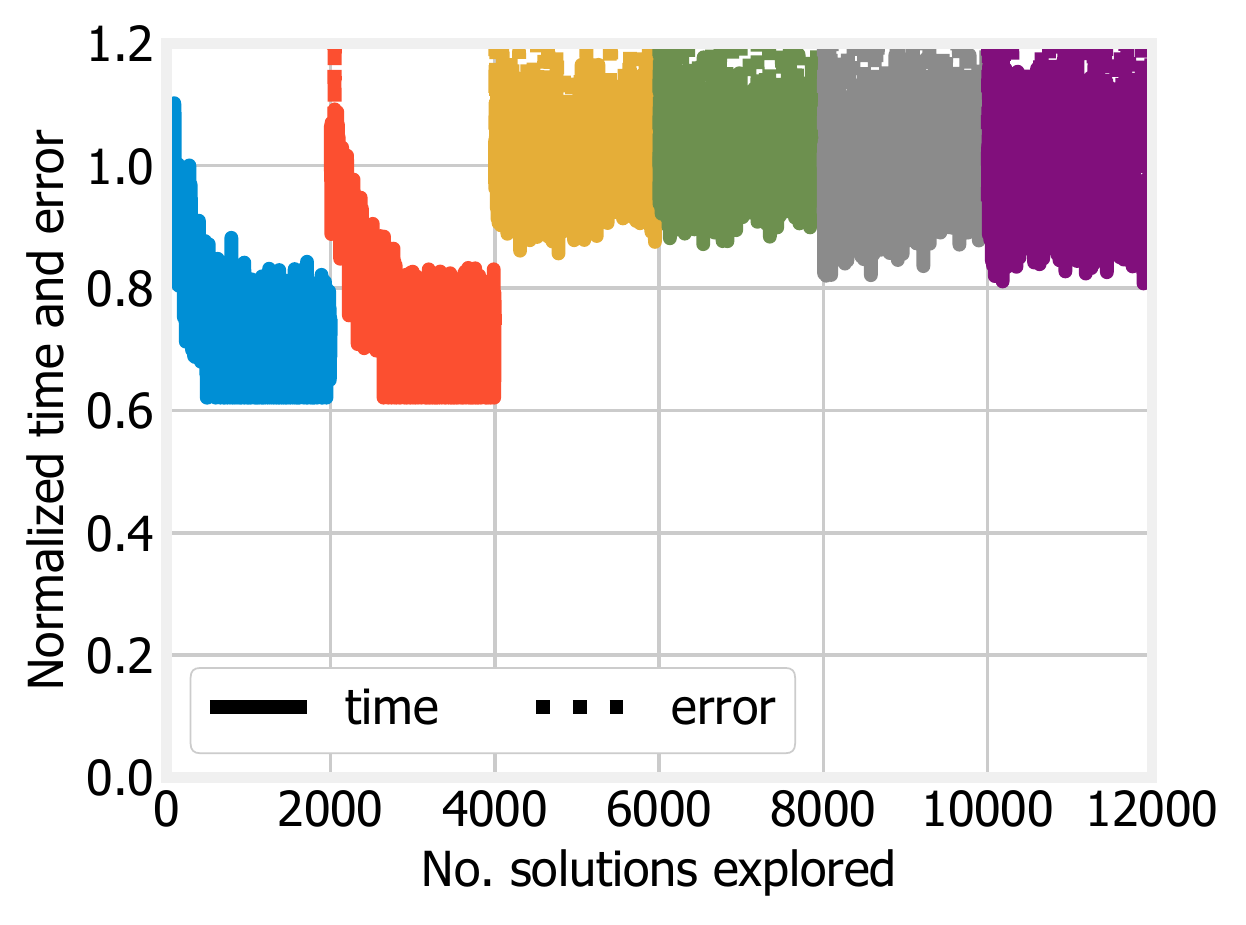}
	\caption{
		{\bf Classification task}: normalized time and error of the solutions examined at each iteration by DoubleClimb (left), Opt-Unif (center), and GA (right), in the basic scenario. Different colors correspond to different values of $d_{\text L}$, as in \Fig{magicsticks}.
		\label{fig:errortime-basic}
	}
	\includegraphics[width=.32\textwidth]{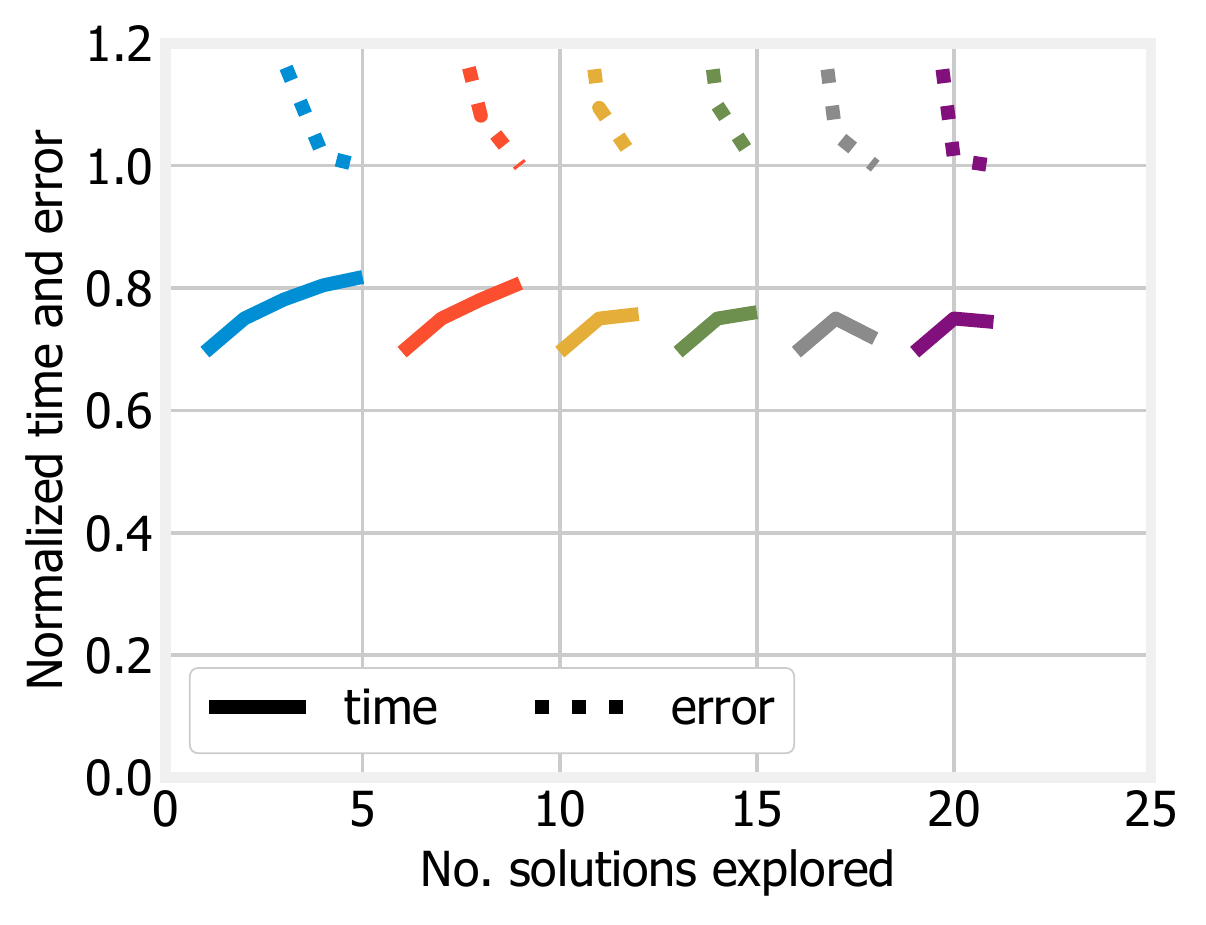}
	\includegraphics[width=.32\textwidth]{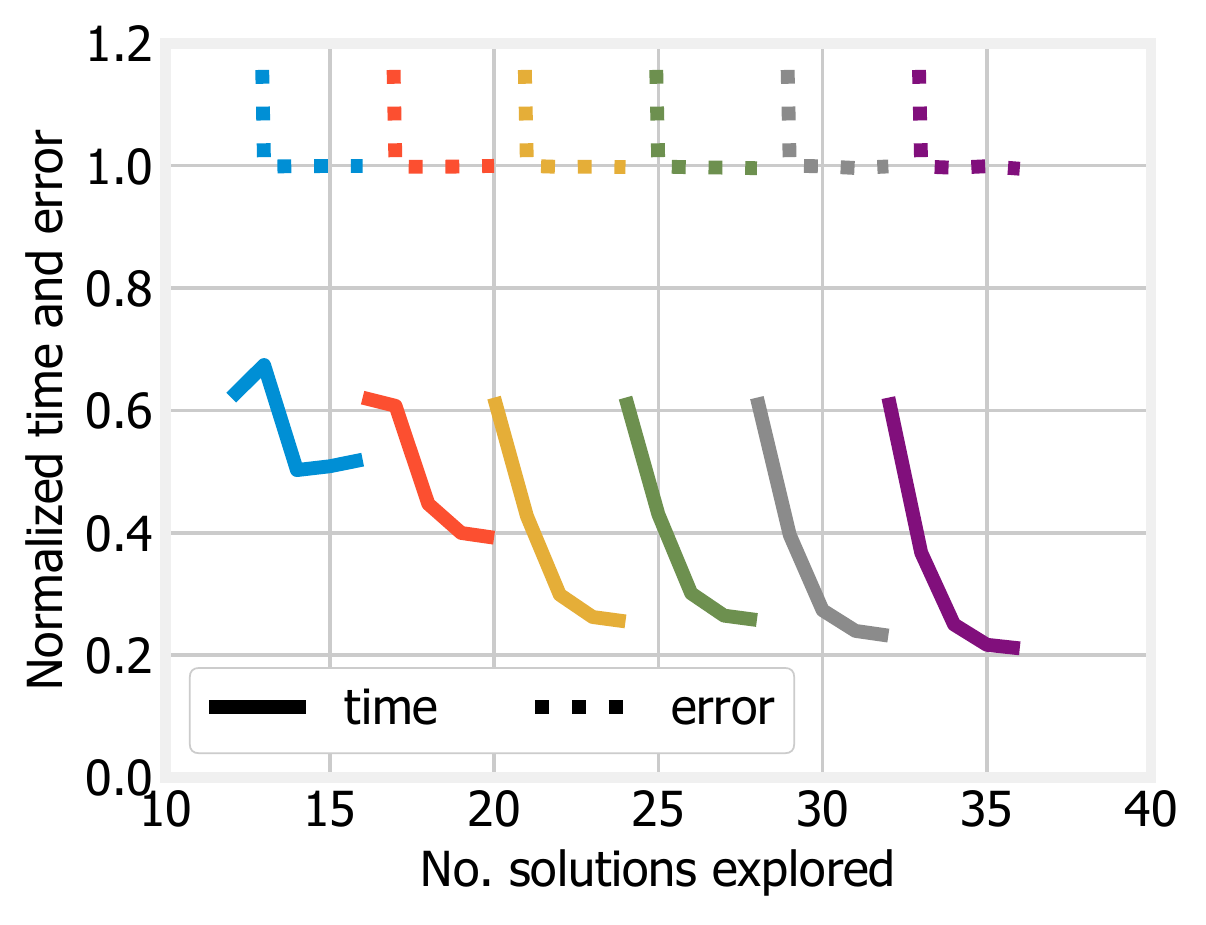}
	\includegraphics[width=.32\textwidth]{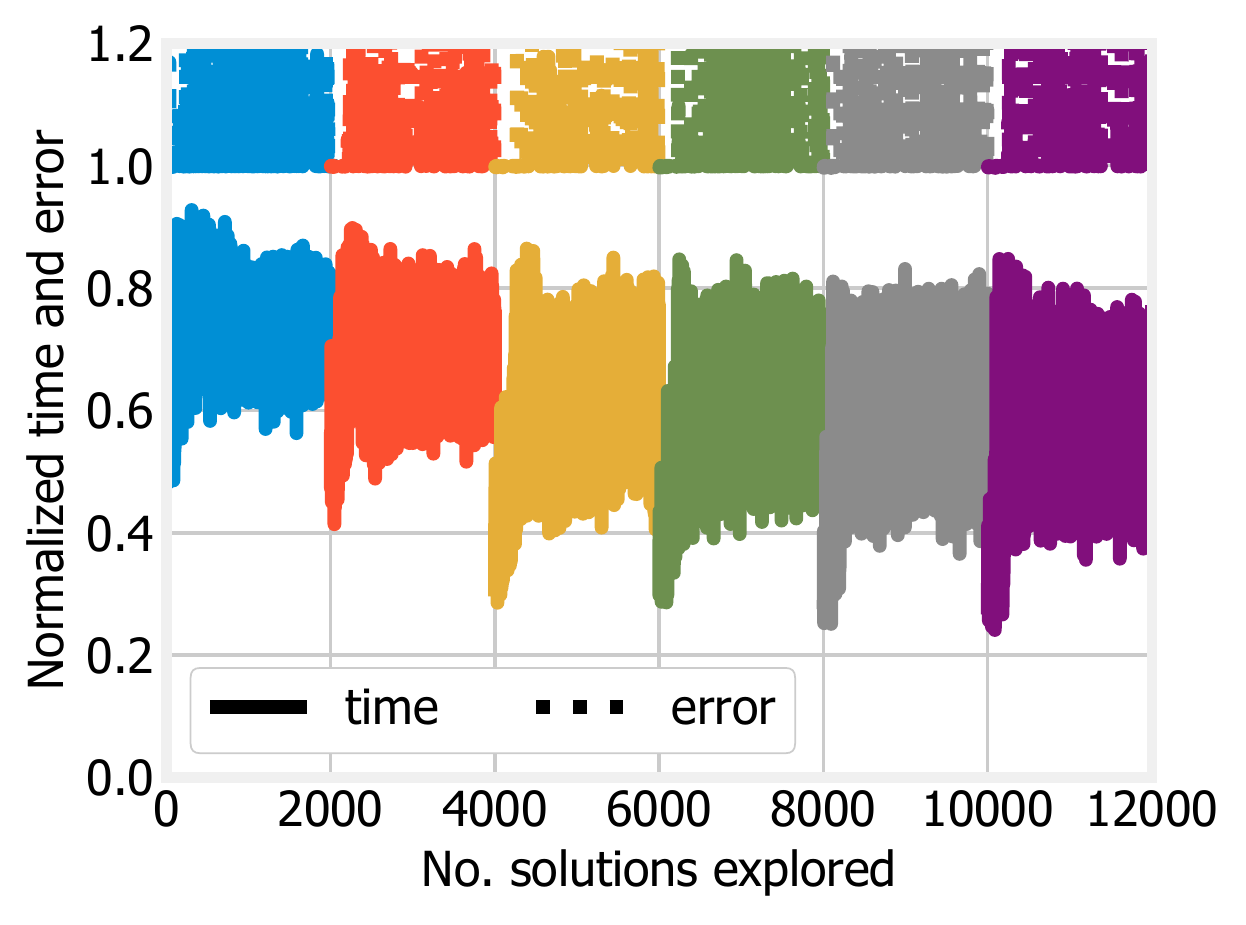}
	\caption{
		{\bf Classification task}: normalized time and error of the solutions examined at each iteration by DoubleClimb (left), Opt-Unif (center), and GA (right), in the rich scenario. Different colors correspond to different values of $d_{\text L}$, as in \Fig{magicsticks}.
		\label{fig:errortime-rich}
	}
	\includegraphics[width=.24\textwidth]{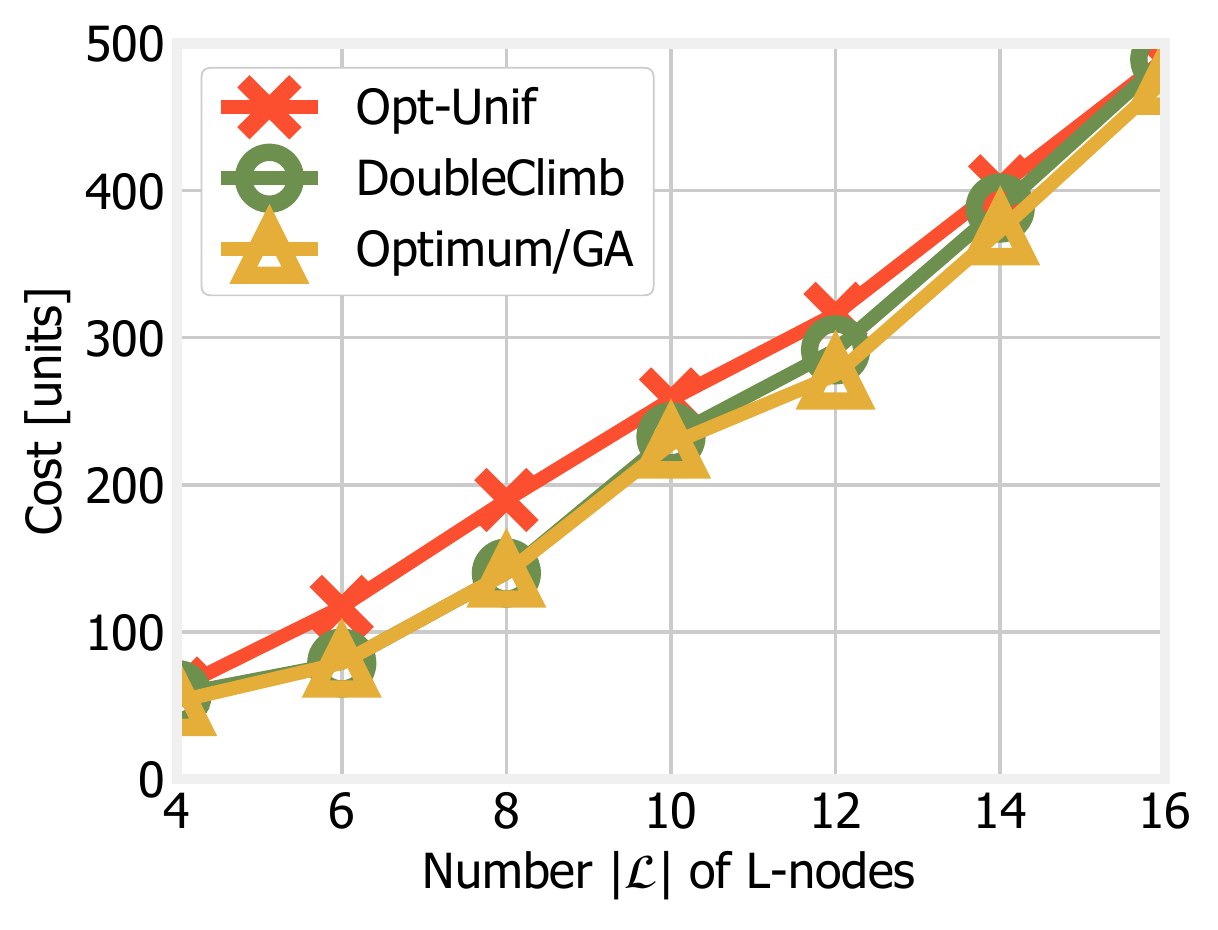}
	\includegraphics[width=.24\textwidth]{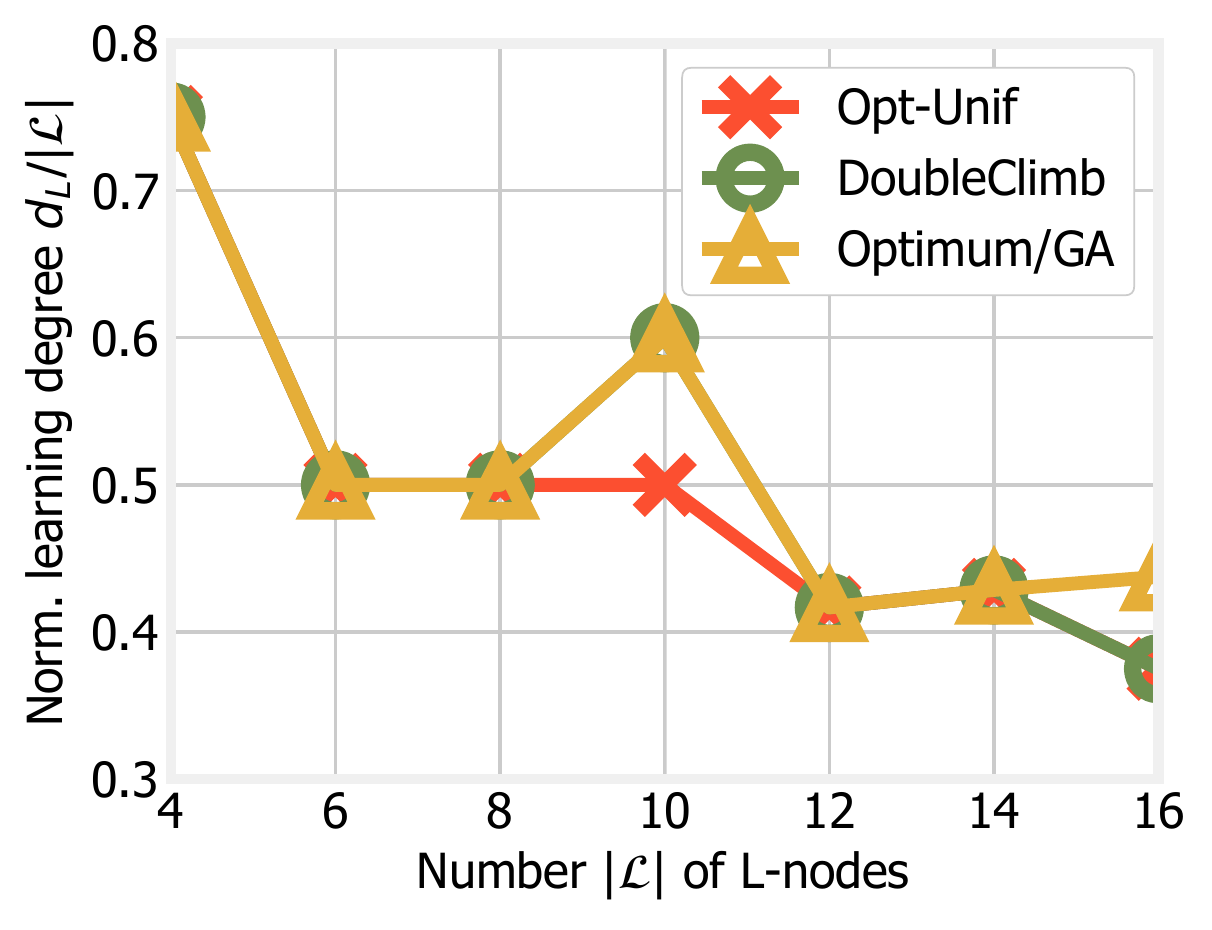}
	\includegraphics[width=.24\textwidth]{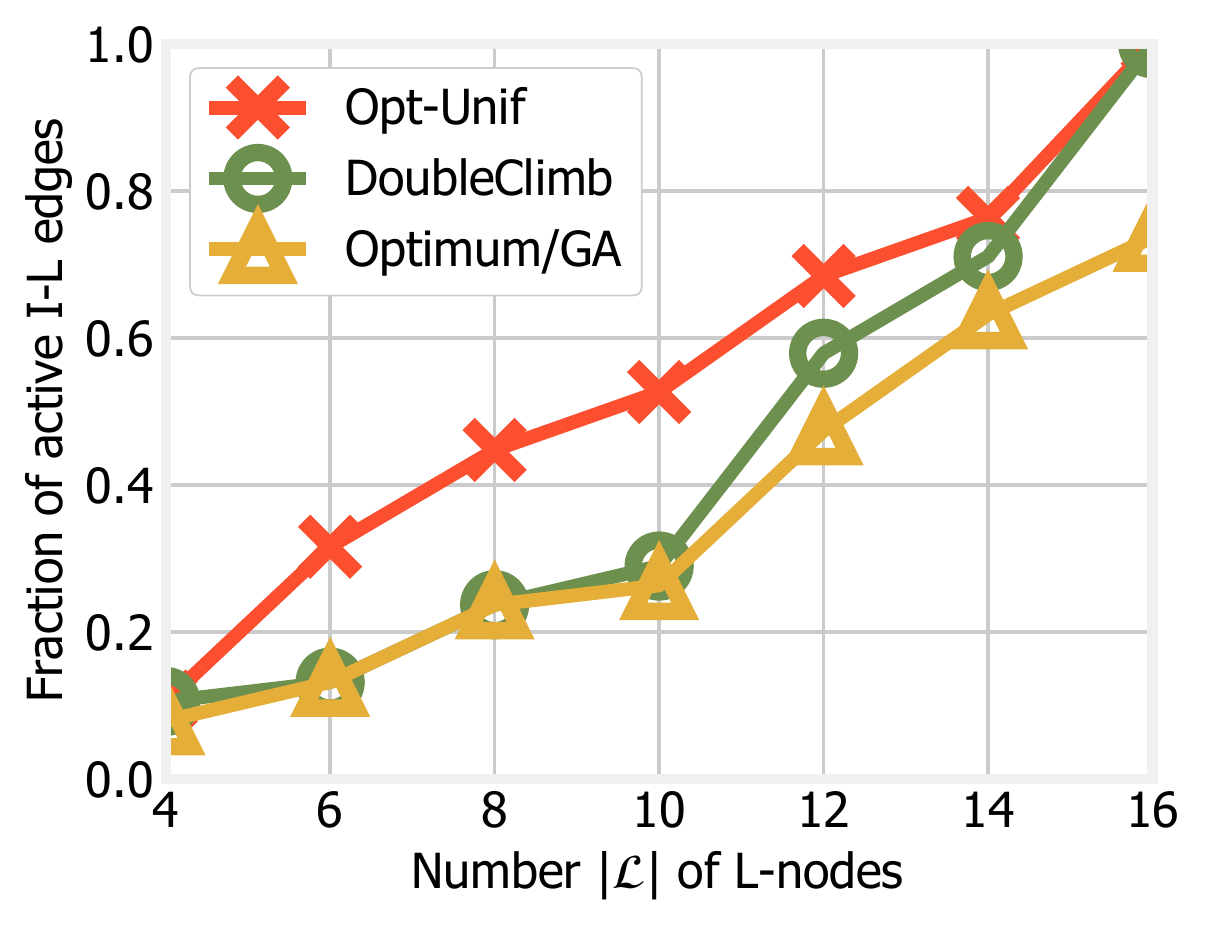}
	\includegraphics[width=.24\textwidth]{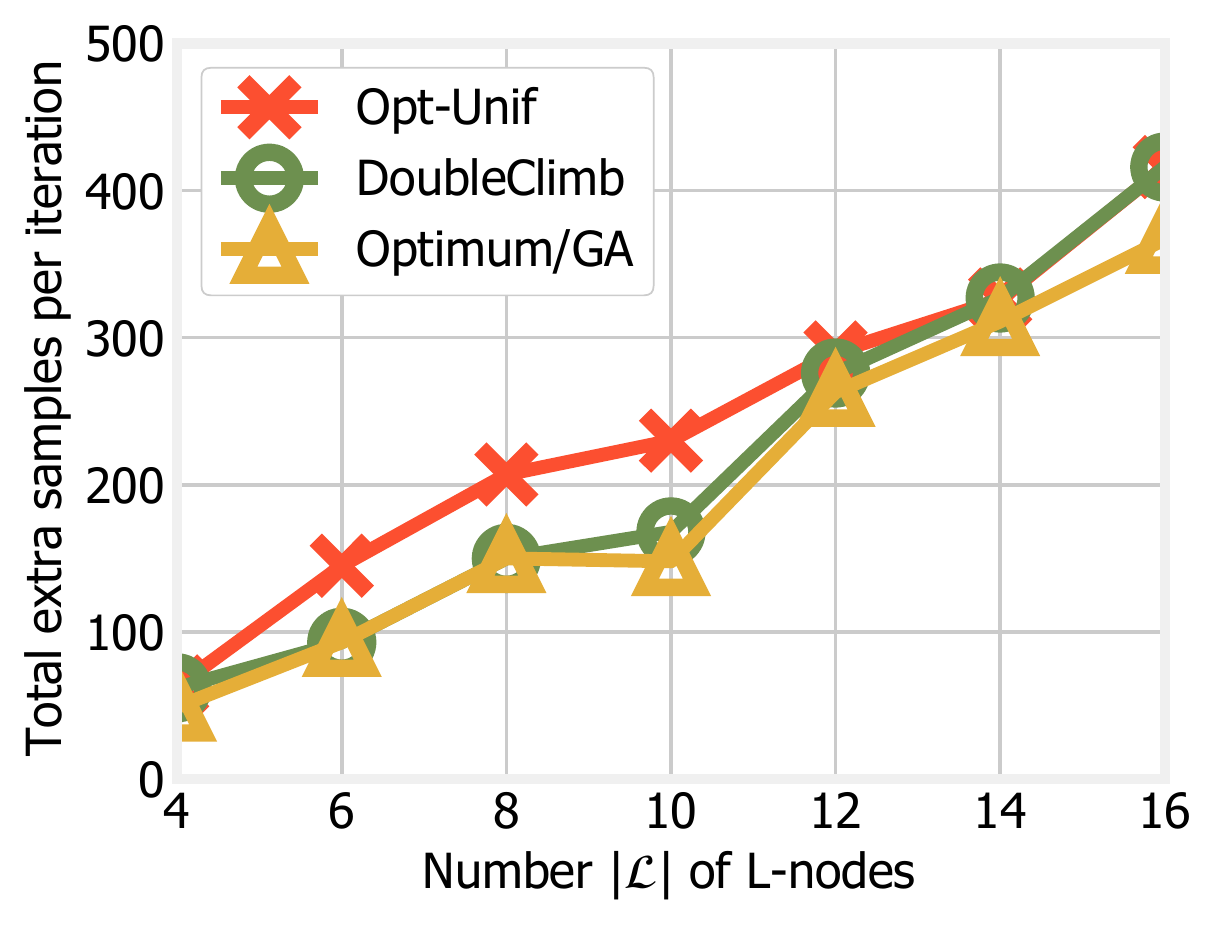}
	\caption{
			{\bf Regression task:} comparison between DoubleClimb, Opt-Unif and the optimum (obtained
			via brute-force) in the basic and rich scenarios, for different
			values of~$|\Lc|$. From left to right: total cost; selected value
			of $d_{\text L}$, normalized (to the
			maximum); fraction of selected I-L edges; total number of extra samples per epoch.
			\label{fig:compare2}
	} 
\end{figure*}

We leverage the parameters for $c_1$--$c_3$ above to compare the performance of DoubleClimb and its alternatives, for both the tasks described earlier.

We begin with the classification task and show, in
the first plot in \Fig{compare}, the cost of DoubleClimb and its
benchmarks, for different  numbers of L-nodes. As expected, the cost
increases with~$|\Lc|$ and decreases in the rich scenario, where the higher quantity of data results in faster convergence. Also, it is clear that
the cost yielded by DoubleClimb is much lower than that of Opt-Unif and matches that of Optimum/GA. 
GA approaches are not, in general, guaranteed to 
yield optimal performance; therefore, we cannot conclude that DoubleClimb makes optimal 
decisions other than for $d_L\leq 6$, when the comparison with brute force was possible. 
However, GA approaches have long been known to be remarkably good at finding optimal 
or near-optimal solutions for combinatorial problems such as the one at hand, 
at the price of long run times, as shown in \Fig{errortime-basic} and \Fig{errortime-rich} next.
Observing that DoubleClimb matches Optimum/GA in all scenarios and for all values of $d_L$ therefore boosts our confidence in the algorithm's effectiveness.

We now look deeper into the decisions made by each strategy. The
second plot in \Fig{compare} depicts the selected value of $d_{\text
  L}$, normalized to $|\Lc|$. Interestingly, such a value is lower in
the rich scenario, confirming our intuition that a tighter cooperation
between L-nodes and more data coming from I-nodes are, to an extent,
alternative solutions to achieve faster learning. DoubleClimb and
Opt-Unif make exactly the same decisions in all cases, which suggests
that the difference in cost shown in the first plot only comes from
the choice of I-L edges. 
Accordingly, the third plot in \Fig{compare}, depicting the fraction of I-L edges selected by each strategy, highlights how DoubleClimb uses substantially fewer edges than Opt-Unif. This highlights how the greater flexibility in the choice of I-L edges is an important asset of our approach, allowing us to beat state-of-the-art alternatives.

The fourth plot in \Fig{compare} shows how DoubleClimb not 
only uses fewer I-L edges, but also chooses the {\em right} ones. The
plot depicts the number of new samples arriving at each epoch
and highlights how, in spite of the substantially smaller number of
selected I-L edges, DoubleClimb obtains a similar number of samples as Opt-Unif. Such an effect is especially evident for the basic scenario, where the number of samples provided by each I-node is smaller.

Comparing the DoubleClimb and Optimum/GA curves, we can observe that in some cases Optimum/GA can activate  
slightly fewer I-L edges than the base scenario, e.g., for $d_L=8$. 
This corresponds to solutions that DoubleClimb is unable to reach due to its hill-climbing nature; 
however, the impact on the overall cost (see the first plot in \Fig{compare}) is negligible. 
Interestingly, DoubleClimb and Optimum/GA make the very same decisions in the rich scenario, 
confirming the somehow counterintuitive notion stated in \Prope{enough}, i.e., 
that, 
the solutions yielded by DoubleClimb tend to be closer to the optimum.

In \Fig{magicsticks}, we seek to better understand how DoubleClimb and
Opt-Unif operate. Every marker in the plots corresponds to one
solution examined by the algorithms; feasible solutions are denoted by
a silver circle, the cheapest of such solutions is denoted by a black
star. 
Note that Opt-Unif explores fewer solutions than DoubleClimb, as 
it is restricted to creating uniform logical topologies. Also, under the rich
scenario it is easier for DoubleClimb to reach a high-quality solution,
hence, the algorithm ends earlier.

The first two plots, representing DoubleClimb in the basic and the
rich scenario, respectively, clearly depict the behavior of
\Alg{doubleclimb}. The algorithm begins with the lowest possible value
of~$d_{\text L}$ and no I-L edges, hence, with a low cost. Then, new
edges are added until either a feasible solution is found, or all I-L
edges are exhausted (as it happens in the first plot, representing the
basic scenario). The double vertical lines in the first two plots
correspond to the triggering of the condition in \Line{check-stop} of \Alg{doubleclimb}; the plots confirm that enforcing such a condition does not result in ignoring cheaper feasible solutions.

The last two plots in \Fig{magicsticks} represent Opt-Unif (again in
the basic and the rich scenario, resp.), and clearly highlight its
differences from DoubleClimb. As mentioned, Opt-Unif tries fewer
solutions; also, multiple feasible solutions are tried out for the
same value of~$d_{\text L}$, since there is no stopping criterion
analogous to the one in \Line{break} in \Alg{doubleclimb}. 
Importantly,  the feasible solutions explored by Opt-Unif are more
costly than those explored by DoubleClimb for the same value
of~$d_{\text L}$, a further confirmation of the importance of a flexible choice of I-L edges.

Last, in \Fig{errortime-basic} and \Fig{errortime-rich}, we examine the error and learning time
associated with each of the solutions examined by DoubleClimb 
and its benchmark solutions, respectively in the basic and rich scenarios. 
Both quantities are normalized to their respective limits, thus 
both lines do not exceed 1 if the corresponding solution is
feasible. It is interesting to note how adding I-L edges (moving from
one solution to the next) affects error and
time. The former (dotted lines) steadily decreases
until its limit is reached, and then stays constant -- recall that the
learning process is interrupted upon reaching~$\epsilon^{\max}$, so
the normalized error never drops substantially below~1. The time 
(solid lines) increases at first, owing to the need to
wait for more I-nodes; then, it decreases due to the fact that
learning can be completed with fewer epochs.
Importantly, both behaviors exactly match those described in \Prope{edges-il} for~$g_1$ and~$g_2$. 
The third plots of both \Fig{errortime-basic} and \Fig{errortime-rich} highlight the behavior of GA approaches, which try multiple different solutions of varying quality and, in the interest of exploration, tend not to abandon low-quality solutions, on the grounds 
that they may mutate into high-quality solutions at some later stage.

The $x$-axis of the plots in \Fig{errortime-basic} and \Fig{errortime-rich} highlight the number of solutions 
being tried out by each of the approaches we study. 
Comparing the first and last plots of each figure, referring to DoubleClimb and Optimal/GA, 
it is easy to observe how the latter examines a number of solutions that is orders of 
magnitude higher than the former. Recalling that, as per \Fig{compare}, the two approaches 
yield a similar performance, it is clear the major efficiency gain brought by DoubleClimb.

Finally, in \Fig{compare2} we examine the performance of DoubleClimb and its 
alternatives for the regression task. 
One can observe that the behavior of  DoubleClimb and the other solutions 
shown  in \Fig{compare2} is consistent with that presented in \Fig{compare}, i.e., 
DoubleClimb can achieve the target learning quality at a lower cost than Opt-Unif, 
by obtaining a similar quantity of data while activating fewer I-L edges. Further, 
the  performance of DoubleClimb matches that of the genetic algorithm. 
This confirms how our model and solution strategy can seamlessly deal 
with different learning tasks. 

\section{Conclusion}
\label{sec:conclusion}

We addressed the problem of 
defining an optimal level of cooperation among network nodes 
performing a supervised learning task.
We  first developed a system model 
accounting for the presence of both  learning nodes and
information nodes interacting with each other. 
Then we formulated the  problem of  choosing  which learning nodes
should cooperate to complete the learning task, and the information
nodes that should provide them with data, as well as 
the number of  epochs to perform. 
Although  being  NP hard, we showed some important properties of our problem, most notably its
submodularity, which allowed us to 
define a solution algorithm that has cubic {\em worst-case} time complexity and is
$1+1/|\Ic|$-competitive, with $\Ic$ being the set of information nodes.
Numerical results also show that our approach closely matches the optimum and outperforms state-of-the-art solutions.

\bibliographystyle{IEEEtran}
\bibliography{biblio}

\begin{thebibliography}{10}
\providecommand{\url}[1]{#1}
\csname url@samestyle\endcsname
\providecommand{\newblock}{\relax}
\providecommand{\bibinfo}[2]{#2}
\providecommand{\BIBentrySTDinterwordspacing}{\spaceskip=0pt\relax}
\providecommand{\BIBentryALTinterwordstretchfactor}{4}
\providecommand{\BIBentryALTinterwordspacing}{\spaceskip=\fontdimen2\font plus
\BIBentryALTinterwordstretchfactor\fontdimen3\font minus
  \fontdimen4\font\relax}
\providecommand{\BIBforeignlanguage}[2]{{%
\expandafter\ifx\csname l@#1\endcsname\relax
\typeout{** WARNING: IEEEtran.bst: No hyphenation pattern has been}%
\typeout{** loaded for the language `#1'. Using the pattern for}%
\typeout{** the default language instead.}%
\else
\language=\csname l@#1\endcsname
\fi
#2}}
\providecommand{\BIBdecl}{\relax}
\BIBdecl

\bibitem{li2014scaling}
M.~Li, D.~G. Andersen, J.~W. Park, A.~J. Smola, A.~Ahmed, V.~Josifovski,
  J.~Long, E.~J. Shekita, and B.-Y. Su, ``Scaling distributed machine learning
  with the parameter server,'' in \emph{USENIX OSDI}, 2014.

\bibitem{pham2018cooperative}
H.~X. Pham, H.~M. La, D.~Feil-Seifer, and A.~Nefian, ``Cooperative and
  distributed reinforcement learning of drones for field coverage,''
  \emph{arXiv preprint arXiv:1803.07250}, 2018.

\bibitem{distributedQlearning}
H.~Y. Ong, K.~Chavez, and A.~Hong, ``Distributed deep q-learning,''
  \emph{CoRR}, 2015.

\bibitem{8340193}
A.~{Nedi\'{c}}, A.~{Olshevsky}, and M.~G. {Rabbat}, ``Network topology and
  communication-computation tradeoffs in decentralized optimization,''
  \emph{Proceedings of the IEEE}, 2018.

\bibitem{wang2019adaptive}
S.~Wang, T.~Tuor, T.~Salonidis, K.~K. Leung, C.~Makaya, T.~He, and K.~Chan,
  ``Adaptive federated learning in resource constrained edge computing
  systems,'' \emph{IEEE Journal on Selected Areas in Communications}, 2019.

\bibitem{zhuo2019federated}
H.~H. Zhuo, W.~Feng, Y.~Lin, Q.~Xu, and Q.~Yang, ``Federated deep reinforcement
  learning,'' \emph{arXiv preprint arXiv:1901.08277}, 2019.

\bibitem{konen2016federatedStrategies}
J.~Konečný, H.~B. McMahan, F.~X. Yu, P.~Richtárik, A.~T. Suresh, and
  D.~Bacon, ``Federated learning: Strategies for improving communication
  efficiency,'' \emph{arXiv preprint arXiv:1610.05492}, 2016.

\bibitem{ETSI-ZSM}
{ETSI}, ``{Zero touch network \& Service Management (ZSM)},''
  \url{https://www.etsi.org/committee/zsm}, online; accessed July 2020.

\bibitem{ETSI-ENI}
------, ``{Experiential Networked Intelligence (ENI)},''
  \url{https://www.etsi.org/committee-activity/eni}, online; accessed July
  2020.

\bibitem{O-RAN}
{Operator Defined Next Generation RAN Architecture and Interfaces}, ``{O-RAN
  Working Group 2: AI/ML workflow description and requirements},'' Tech. Rep.
  O-RAN.WG2.AIML-v01.01, online; accessed July 2020.

\bibitem{xiao2020towards}
Y.~Xiao, G.~Shi, Y.~Li, W.~Saad, and H.~V. Poor, ``Towards self-learning edge
  intelligence in 6g,'' \emph{IEEE Communications Magazine}, 2020.

\bibitem{ETSI-36}
{ETSI}, ``{MEC Working Item 36, MEC in resource constrained terminals, fixed or
  mobile},'' \url{https://portal.etsi.org/webapp/WorkProgram/}, online;
  accessed July 2020.

\bibitem{kadav2016asap}
A.~Kadav and E.~Kruus, ``Asap: asynchronous approximate data-parallel
  computation,'' \emph{arXiv preprint arXiv:1612.08608}, 2016.

\bibitem{li2018near}
S.~Li, S.~M.~M. Kalan, A.~S. Avestimehr, and M.~Soltanolkotabi, ``Near-optimal
  straggler mitigation for distributed gradient methods,'' in \emph{IEEE
  IPDPSW}, 2018.

\bibitem{neglia}
G.~{Neglia}, G.~{Calbi}, D.~{Towsley}, and G.~{Vardoyan}, ``The role of network
  topology for distributed machine learning,'' in \emph{IEEE INFOCOM}, 2019.

\bibitem{levine2020offline}
S.~Levine, A.~Kumar, G.~Tucker, and J.~Fu, ``Offline reinforcement learning:
  Tutorial, review, and perspectives on open problems,'' \emph{arXiv preprint
  arXiv:2005.01643}, 2020.

\bibitem{alaa}
A.~A. Abdellatif, C.~F. Chiasserini, and F.~Malandrino, ``Active learning-based
  classification in automated connected vehicles,'' in \emph{IEEE INFOCOM
  PERSIST-IoT Workshop}, 2020.

\bibitem{8600752}
K.~{Yang}, J.~{Ren}, Y.~{Zhu}, and W.~{Zhang}, ``Active learning for wireless
  iot intrusion detection,'' \emph{IEEE Wireless Communications}, 2018.

\bibitem{chen2018communication}
T.~Chen, K.~Zhang, G.~B. Giannakis, and T.~Ba{\c{s}}ar,
  ``Communication-efficient distributed reinforcement learning,'' \emph{arXiv
  preprint arXiv:1812.03239}, 2018.

\bibitem{li2019accelerating}
Y.~Li, I.-J. Liu, Y.~Yuan, D.~Chen, A.~Schwing, and J.~Huang, ``Accelerating
  distributed reinforcement learning with in-switch computing,'' in
  \emph{ISCA}, 2019.

\bibitem{konen2015federatedOptimization}
J.~Konečný, B.~McMahan, and D.~Ramage, ``Federated optimization: Distributed
  optimization beyond the datacenter,'' \emph{arXiv preprint arXiv:1511.03575},
  2015.

\bibitem{shamir2013stochastic}
O.~Shamir and T.~Zhang, ``Stochastic gradient descent for non-smooth
  optimization: Convergence results and optimal averaging schemes,'' in
  \emph{International conference on machine learning}, 2013.

\bibitem{hard2018federated}
A.~Hard, K.~Rao, R.~Mathews, S.~Ramaswamy, F.~Beaufays, S.~Augenstein,
  H.~Eichner, C.~Kiddon, and D.~Ramage, ``Federated learning for mobile
  keyboard prediction,'' \emph{arXiv preprint arXiv:1811.03604}, 2018.

\bibitem{MQTT}
{OASIS Standard}, ``{MQTT Version 5.0, Mar. 2019},''
  \url{https://docs.oasis-open.org/mqtt/mqtt/v5.0/mqtt-v5.0.html}, online;
  accessed July 2020.

\bibitem{zenoh}
``{zenoh: Zero Overhead Pub/sub, Store/Query and Compute},''
  \url{http://zenoh.io}, online; accessed July 2020.

\bibitem{TS23.501}
{3GPP}, ``{TS23.501, System architecture for the 5G System (5GS), Rel.\,15},''
  \url{https://portal.3gpp.org/desktopmodules/Specifications/SpecificationDetails.aspx?specificationId=3144},
  online; accessed July 2020.

\bibitem{yang2019federated}
Q.~Yang, Y.~Liu, T.~Chen, and Y.~Tong, ``Federated machine learning: Concept
  and applications,'' \emph{ACM Transactions on Intelligent Systems and
  Technology}, 2019.

\bibitem{nagelkerke1991note}
N.~J. Nagelkerke \emph{et~al.}, ``A note on a general definition of the
  coefficient of determination,'' \emph{Biometrika}, 1991.

\bibitem{hestness2017deep}
J.~Hestness, S.~Narang, N.~Ardalani, G.~Diamos, H.~Jun, H.~Kianinejad,
  M.~Patwary, M.~Ali, Y.~Yang, and Y.~Zhou, ``Deep learning scaling is
  predictable, empirically,'' \emph{arXiv preprint arXiv:1712.00409}, 2017.

\bibitem{sun2017revisiting}
C.~Sun, A.~Shrivastava, S.~Singh, and A.~Gupta, ``Revisiting unreasonable
  effectiveness of data in deep learning era,'' in \emph{IEEE ICCV}, 2017.

\bibitem{linjordet2019impact}
T.~Linjordet and K.~Balog, ``Impact of training dataset size on neural answer
  selection models,'' in \emph{European Conference on Information Retrieval},
  2019.

\bibitem{perlich2003tree}
C.~Perlich, F.~Provost, and J.~S. Simonoff, ``Tree induction vs. logistic
  regression: A learning-curve analysis,'' \emph{Journal of Machine Learning
  Research}, 2003.

\bibitem{serpen2014complexity}
G.~Serpen and Z.~Gao, ``Complexity analysis of multilayer perceptron neural
  network embedded into a wireless sensor network,'' \emph{Procedia Computer
  Science}, 2014.

\bibitem{multinomial}
D.~Bolton, ``The multinomial theorem,'' \emph{The Mathematical Gazette}, pp.
  336--342, 1968.

\bibitem{woeginger2003exact}
G.~J. Woeginger, ``Exact algorithms for {NP}-hard problems: {A} survey,'' in
  \emph{Combinatorial optimization—eureka, you shrink!}\hskip 1em plus 0.5em
  minus 0.4em\relax Springer, 2003.

\bibitem{lovasz1983submodular}
L.~Lov{\'a}sz, ``Submodular functions and convexity,'' in \emph{Mathematical
  Programming The State of the Art}.\hskip 1em plus 0.5em minus 0.4em\relax
  Springer, 1983.

\bibitem{conforti1984submodular}
M.~Conforti and G.~Cornu{\'e}jols, ``Submodular set functions, matroids and the
  greedy algorithm: tight worst-case bounds and some generalizations of the
  rado-edmonds theorem,'' \emph{Discrete applied mathematics}, 1984.

\bibitem{vu2008random}
V.~Vu, ``Random discrete matrices,'' in \emph{Horizons of combinatorics}.\hskip
  1em plus 0.5em minus 0.4em\relax Springer, 2008.

\bibitem{tikhomirov2019spectral}
K.~Tikhomirov and P.~Youssef, ``The spectral gap of dense random regular
  graphs,'' \emph{The Annals of Probability}, 2019.

\bibitem{iyer2013submodular}
R.~K. Iyer and J.~A. Bilmes, ``Submodular optimization with submodular cover
  and submodular knapsack constraints,'' in \emph{Advances in Neural
  Information Processing Systems}, 2013.

\bibitem{valerio2018energy}
L.~Valerio, M.~Conti, and A.~Passarella, ``Energy efficient distributed
  analytics at the edge of the network for iot environments,'' \emph{Elsevier
  Pervasive and Mobile Computing}, 2018.

\bibitem{ye2018machine}
H.~Ye, L.~Liang, G.~Y. Li, J.~Kim, L.~Lu, and M.~Wu, ``Machine learning for
  vehicular networks: Recent advances and application examples,'' \emph{IEEE
  Vehicular Technology Magazine}, 2018.

\bibitem{diro2018distributed}
A.~A. Diro and N.~Chilamkurti, ``Distributed attack detection scheme using deep
  learning approach for internet of things,'' \emph{Future Generation Computer
  Systems}, 2018.

\bibitem{CrosshaulD12}
{5G-Crosshaul}, ``{D1.2: Final 5G-Crosshaul system design and economic
  analysis},'' December 2017.

\bibitem{digits}
L.~Deng, ``The mnist database of handwritten digit images for machine learning
  research,'' \emph{IEEE Signal Processing Magazine}, 2012.

\bibitem{ituchallenge}
{ITU}, ``{AI/ML in 5G Challenge 2020},''
  \url{https://www.itu.int/en/ITU-T/AI/challenge/2020/}, online; accessed
  November 2020.

\end{thebibliography}
\begin{IEEEbiographynophoto}{Francesco Malandrino} (M'09, SM'19) earned his Ph.D. degree from Politecnico di Torino in 2012 and is now a researcher at the National Research Council of Italy (CNR-IEIIT). His research interests include the architecture and management of wireless, cellular, and vehicular networks.
\end{IEEEbiographynophoto}
\begin{IEEEbiographynophoto}{Carla Fabiana Chiasserini} (F'18) worked as a visiting researcher at 
UCSD, and as a Visiting Professor at Monash University in 2012 and 2016 and at TUB in 2021 and 2022.  
She is currently a  Professor  at
Politecnico di Torino and the EiC of Computer Communications.  
\end{IEEEbiographynophoto}
\begin{IEEEbiographynophoto} 
{Nuria Molner} obtained her Ph.D. in Telematics Engineering from University
Carlos III of Madrid in 2021. Currently, she is a researcher at Universitat
Polit\`ecnica de Val\`encia (iTEAM-UPV).
\end{IEEEbiographynophoto}
\begin{IEEEbiographynophoto}
{Antonio de la Oliva} received his M.Sc. degree in 2004 and his Ph.D. degree in 2008. He is an associate professor at Universidad Carlos III de Madrid.
\end{IEEEbiographynophoto}

\end{document}